\documentclass[singlecolumn,showpacs,preprintnumbers,amsmath,amssymb,prb]{revtex4}

\usepackage{graphicx}
\usepackage{dcolumn}

\begin{document}
\title{Scaling analysis of Kondo screening cloud in
a mesoscopic ring with \\ an embedded quantum dot}

\author{Ryosuke
Yoshii\thanks{E-mail address: ryoshii@phys.keio.ac.jp}
and Mikio Eto}
\affiliation{Faculty of Science and Technology, Keio University,
3-14-1 Hiyoshi, Kohoku-ku, Yokohama 223-8522, Japan}

\date{today}

\begin{abstract}
{The Kondo effect is theoretically studied in a quantum dot 
embedded in a mesoscopic ring. The ring is connected to two
external leads, which enables the transport measurement.
Using the ``poor man's'' scaling method, we obtain analytical
expressions of the Kondo temperature $T_\mathrm{K}$
as a function of the Aharonov-Bohm phase $\phi$ by
the magnetic flux penetrating the ring.
In this Kondo problem, there are two characteristic
lengths. One is the screening length of the charge
fluctuation,
$L_\mathrm{c}=\hbar v_{\rm F}/ |\epsilon_0|$,
where $v_{\rm F}$ is the Fermi velocity and
$\epsilon_0$ is the energy level
in the quantum dot. The other is the screening length of
spin fluctuation, i.e., size of Kondo screening cloud,
$L_\mathrm{K}=\hbar v_{\rm F}/ T_\mathrm{K}$.
We obtain different expressions of $T_\mathrm{K}(\phi)$
for (i) $L_{\rm c} \ll L_{\rm K} \ll L$,
(ii) $L_{\rm c} \ll L \ll L_{\rm K}$,
and (iii) $L \ll L_{\rm c} \ll L_{\rm K}$, where $L$ is the
size of the ring. $T_\mathrm{K}$ is markedly modulated
by $\phi$ in cases (ii) and (iii), whereas it hardly depends
on $\phi$ in case (i).
We also derive logarithmic corrections to the conductance at temperature 
$T\gg T_{\mathrm{K}}$ and an analytical expression of the conductance at 
$T\ll T_{\mathrm{K}}$, on the basis of the scaling analysis.
}
\end{abstract}

\maketitle

\section{Introduction}

The Kondo effect is one of the most important and fundamental
problems in condensed matter physics.\cite{Kondo,Hewson}
When a localized spin
contacts with the electron Fermi sea, the many-body state of
spin-singlet is locally formed at temperatures lower than
the Kondo temperature $T_\mathrm{K}$.
An open problem in the Kondo physics is the
observation of the many-body state, so-called Kondo
screening cloud.
The size of the screening cloud is evaluated as
\begin{equation}
L_{\mathrm{K}}= \hbar v_{\mathrm{F}}/T_{\mathrm{K}},
\label{eq:LK}
\end{equation}
where $v_{\mathrm{F}}$ is the Fermi velocity.
There have been several theoretical proposals for the
observation of $L_{\mathrm{K}}$,\cite{Affleck0}
e.g., the Knight shift as a function of the distance from a
magnetic impurity in metal,\cite{Sorensen,Barzykin1,Barzykin2}
ring-size dependence of the persistent current in an
isolated ring with an embedded quantum
dot,\cite{Affleck,Simon1,Sorensen2,Affleck2}
the Friedel oscillation around a magnetic impurity
in metal,\cite{Affleck3} and the spin-spin correlation
function.\cite{Borda,Holzner}
In the present paper, we theoretically examine the
Kondo effect in a quantum dot embedded in a mesoscopic
ring to elucidate the effects of the formation of Kondo
screening cloud on the physical properties, based on the
scaling analysis.

The Kondo effect in quantum dots has been intensively studied
in a conventional geometry in which a quantum dot connected
to two external leads. At $T \gg T_{\mathrm{K}}$,
the current through the quantum dot
shows a peak structure, so-called Coulomb oscillation,
when the electrostatic potential in the dot is
changed by the gate voltage.
Between the current peaks, the number of electrons is
almost fixed by the Coulomb blockade. With an odd number
of electrons, the tunnel coupling between a localized
spin $1/2$ in the dot and conduction electrons in
the leads results in the Kondo effect at
$T < T_{\mathrm{K}}$. The resonant tunneling of conduction
electrons through the many-body Kondo state enhances the
conductance to of the order of $2e^2/h$ at
$T \ll T_{\mathrm{K}}$.\cite{Goldhaber-Gordon,Cronenwett,Wiel}
Various aspects of the Kondo effect has been elucidated in
the quantum dot owing to its artificial tunability and
flexibility, e.g., an enhanced Kondo effect with an even
number of electrons at the spin-singlet-triplet
degeneracy,\cite{Sasaki1} the SU(4) Kondo effect with
$S=1/2$ and orbital degeneracy,\cite{Sasaki2}
bonding and antibonding states between the Kondo resonant
levels in coupled quantum dots,\cite{Aono,Jeong} and
Kondo effect in a quantum dot coupled to ferromagnetic
leads.\cite{Martinek,Pasupathy}

Mesoscopic rings with an embedded quantum dot are also
fabricated and being studied. The rings are connected to
source and drain leads, which enables to examine the
coherent transport through the Aharonov-Bohm (AB) effect.
Using the so-called AB interferometers,
the transmission phase of an electron passing through
a quantum dot was measured in the
absence\cite{Yacoby,Schuster} or presence of
the Kondo effect.\cite{Gerland,Ji}
Without the Kondo effect, the Fano resonance of asymmetric
shape with a peak and a dip is observed as a function of the
gate voltage, which stems from the interference between
a discrete level in the quantum dot and continuum spectrum
in the ring.\cite{Kobayashi}
In the Kondo regime, the one-body interference effect and
many-body Kondo effect coexist, which modifies the Fano
resonant shape with phase locking at $\pi/2$ due to the
Kondo many-body resonance.
This Fano-Kondo effect was studied by several theoretical
groups using a minimal model with a single energy level
$\epsilon_0$ in the quantum dot and in the small limit of
ring size,\cite{Bulka,Hofstetter,Konik,Maruyama}
e.g., using the equation-of-motion method with the Green
function,\cite{Bulka} the numerical renormalization group
method,\cite{Hofstetter} the exact solution by the Bethe
ansatz,\cite{Konik} and the density-matrix renormalization
group method.\cite{Maruyama} The character of the Fano-Kondo
effect was reported by recent experiment.\cite{Katsumoto}
In the present paper,
we concentrate on the Kondo regime in this system.

In the AB interferometer in the Kondo regime, the Kondo
screening cloud should be affected by the AB interference
effect if the screening cloud is larger than the
size of the ring. Although the interference effect on the
value of $T_{\mathrm{K}}$ was studied by some
groups,\cite{Simon2,Malecki} the magnetic-flux
dependence of $T_{\mathrm{K}}$ is still controversial.
In our previous work,\cite{Yoshii}
we studied this Kondo problem in the small limit of
ring size, using the ``poor man's'' scaling
method.\cite{Anderson}
The scaling method is suitable for revealing the
Kondo physics in this system and obtaining analytical
expressions of $T_{\mathrm{K}}$ and conductance.
Our calculation method is as follows.
First, we construct an equivalent model in
which a quantum dot is coupled to a single lead.
The AB interference effect is involved in the magnetic-flux
dependence of the density of states in the lead. Next,
the two-stage scaling method\cite{Haldane} is applied
to the reduced model.
On the first stage of scaling, we renormalize
the energy level in the quantum dot by taking into account
the charge fluctuation in the dot. On the second stage,
the spin fluctuation is considered.
The Kondo temperature $T_\mathrm{K}$ is evaluated as a
function of magnetic flux penetrating the ring.
We showed that $T_\mathrm{K}$ is significantly modulated
by the magnetic flux.
The scaling method also yields the logarithmic corrections
to the conductance at temperatures $T \gg T_\mathrm{K}$
and an analytical expression of the conductance at
$T \ll T_\mathrm{K}$.

In the present work, we apply our calculation method to
the Kondo effect in the AB interferometer with finite
size of the ring.
There are two characteristic lengths in this problem.
One is the screening length of the charge fluctuation,
\begin{equation}
L_\mathrm{c}=\hbar v_{\rm F}/ |\epsilon_0|,
\label{eq:Lc}
\end{equation}
where $\epsilon_0$ is the energy level in the quantum dot.
The other is the screening length of
spin fluctuation, i.e., size of the Kondo screening cloud,
$L_\mathrm{K}$ in Eq.\ (\ref{eq:LK}).
We obtain analytical expressions of $T_\mathrm{K}(\phi)$
for (i) $L_{\rm c} \ll L_{\rm K} \ll L$, 
(ii) $L_{\rm c} \ll L \ll L_{\rm K}$,
and (iii) $L \ll L_{\rm c} \ll L_{\rm K}$, where $L$ is the
size of the ring. $T_\mathrm{K}$ is markedly modulated
by $\phi$ in cases (ii) and (iii), whereas it hardly depends
on $\phi$ in case (i). This result clearly indicates that the
Kondo screening cloud is modified by the AB interference
effect when the ring size is smaller than $L_{\rm K}$.
The conductance in analytical forms is also given for
$T \gg T_\mathrm{K}$ and $T \ll T_\mathrm{K}$.

In our model, we consider a single energy level $\epsilon_0$
in the quantum dot.
Regarding the electron-electron interaction $U$ in the dot,
two situations are examined. One is the case of
$U \rightarrow \infty$ and the other is in the vicinity of
electron-hole symmetry, $-\epsilon_0 \simeq \epsilon_0+U$.
The latter case corresponds
to the midpoint between the current peaks in the Coulomb
blockade region. With approaching one of the current peaks,
the situation becomes similar to the case of
$U \rightarrow \infty$. Hence the two situations may be
realized by changing the gate voltage in experiments.

Note that we can accurately evaluate the exponential part
of $T_\mathrm{K}$ by the ``poor man's'' scaling
method,\cite{Hewson} in extreme cases of $L_{\rm K} \ll L$,
$L \ll L_{\rm K}$, etc. The conductance $G$ is properly estimated
only for $T \gg T_\mathrm{K}$ and $T \ll T_\mathrm{K}$.
Accurate evaluations of $T_\mathrm{K}$ and $G$ in intermediate
regimes require the calculations using the numerical
renormalization group method,
which is beyond the scope of the present paper.
We believe, however, that analytical expressions of $T_\mathrm{K}$
and $G$ that we obtain in limited situations
will importantly contribute to understanding the properties of
the Kondo screening cloud in mesoscopic rings.

The organization of the present paper is as follows.
In Sec.\ II, we describe our model for a mesoscopic ring with
an embedded quantum dot. From the original model, we construct
an equivalent model in which a quantum dot is connected to a
single lead.
In Sec.\ III, we perform the two-stage scaling analysis using
the reduced model, in the case of $U \rightarrow \infty$.
Two characteristic lengths, $L_\mathrm{c}$ and
$L_\mathrm{K}$, are naturally derived from the calculations.
We obtain the analytical expressions of the
renormalized energy level in the quantum dot and Kondo
temperature, in the above-mentioned three situations
concerning the ring size $L$.
Section IV is devoted to the scaling analysis in the vicinity
of electron-hole symmetry.
In Sec.\ V, we evaluate the logarithmic corrections to the
conductance at $T \gg T_\mathrm{K}$ and obtain an analytical
expression of the conductance at $T \ll T_\mathrm{K}$,
on the basis of the scaling analysis.
Conclusions and remarks are given in Sec.\ VI.

In Appendix A, we illustrate the two-stage scaling analysis
of the Kondo effect by applying it to the conventional system of
a quantum dot connected to two leads, depicted in Fig.\ 1(b).
In Appendix B, we summerize our previous study on a ring
system with an embedded quantum dot in the small limit of
ring size.\cite{Yoshii} The same model was examined by
Malecki and Affleck,\cite{Malecki} but their results are slightly
different from ours. The reason for the discrepancy is elucidated.

\section{MODEL AND METHODS}

In this section, we present our model for a mesoscopic ring
with an embedded quantum dot. The ring is connected to
source and drain leads. From this model, we construct an
equivalent model in which a quantum dot is connected to
a single lead. The reduced model is more tractable than
the original model by various calculation methods
for the Kondo effect.

\begin{figure}
\includegraphics[width=15pc]{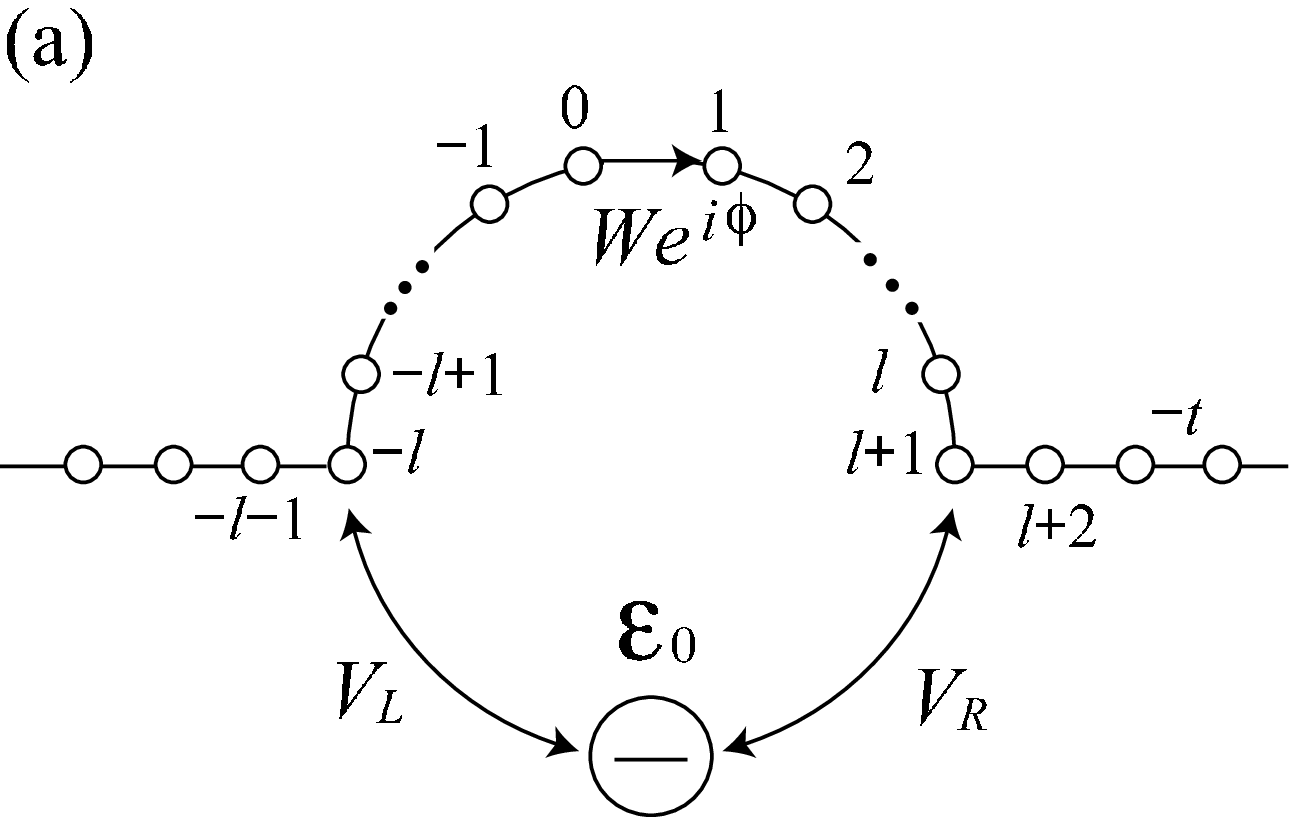}\hspace{1pc}

\includegraphics[width=15pc]{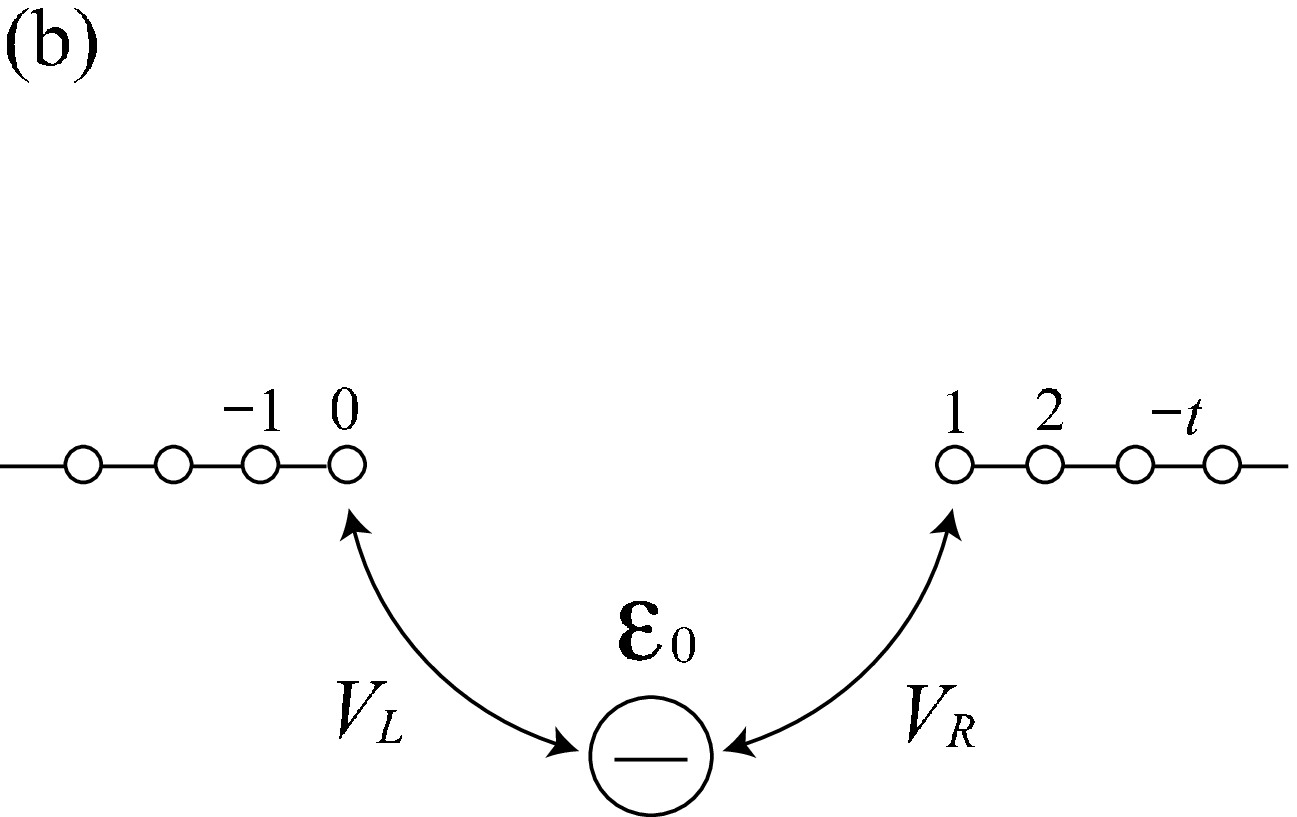}\hspace{1pc}
\caption{\noindent
(a) Model for a mesoscopic ring with an embedded
quantum dot. A quantum dot with single energy level
$\epsilon_0$ is connected to two external leads by
tunnel couplings, $V_L$ and $V_R$.
There is a barrier with tunnel coupling $W$
on an arm of the ring which directly connects the two
leads (reference arm). The ring and two leads are
described by a one-dimensional tight-binding model.
The magnetic flux penetrating the ring is represented by
the AB phase $\phi$ at the tunnel barrier.
(b) Model for a quantum dot coupled to two leads without
the reference arm.
}
\end{figure}

\subsection{Model Hamiltonian}

Our model is depicted in Fig.\ 1(a).
A quantum dot with a single energy level, $\epsilon_0$, is
connected to two external leads by tunnel couplings,
$V_L$ and $V_R$. A barrier with tunnel coupling $W$
is put on an arm of the ring which directly connects the
two leads (reference arm). The reference arm and two leads
are represented by a one-dimensional tight-binding model
with transfer integral $-t$ and lattice constant $a$.
The ring size is defined as $L=(2l+1)a$, where
$2l$ is the number of sites on the reference arm.

When a magnetic flux $\Phi$ penetrates the ring, the
AB phase is given by $\phi=2\pi\Phi/\Phi_0$ with the
flux quantum $\Phi_0=h/e$. The AB interference
effect is considered as the AB phase at the tunnel barrier
without the loss of generality.\cite{com0}
The Hamiltonian of the system reads
\begin{eqnarray}
H^{(0)} &=& H_\mathrm{dot}+H_\mathrm{leads+ring}
+H_\mathrm{T},
\label{eq:orig}
\\
H_\mathrm{dot} &=&
\sum_{\sigma=\uparrow,\downarrow}
\epsilon_0 d^\dagger_\sigma d_\sigma +
U \hat n_{\uparrow} \hat n_{\downarrow},
\\ 
H_\mathrm{leads+ring} &=&
\sum_{i \ne 0}\sum_{\sigma}
(-ta^\dagger_{i+1,\sigma}a_{i,\sigma}+\mathrm{h.c}.)
\nonumber \\
& & +
\sum_{\sigma}(We^{i\phi} a^\dagger_{1,\sigma} a_{0,\sigma}+
\mathrm{h.c}.),
\label{eq:leads+ring}
\\
H_\mathrm{T} &=&
\sum_{\sigma}(V_L d^\dagger_\sigma a_{-l,\sigma}+
V_R d^\dagger_\sigma a_{l+1,\sigma}+\mathrm{h.c}.),
\label{eq:origend}
\end{eqnarray}
where $d_{\sigma}^{\dagger}$ and $d_{\sigma}$ are creation and
annihilation operators, respectively, of an electron
in the quantum dot with spin $\sigma$.
$a_{i,\sigma}^{\dagger}$ and $a_{i,\sigma}$ are
those at site $i$ with spin $\sigma$ in the leads or ring. 
$U$ is the electron-electron interaction in the quantum dot.
$\hat n_\sigma=d^\dagger_\sigma d_\sigma$ is the number
operator in the dot with spin $\sigma$.

In the leads, the energy dispersion is linearlized
around the Fermi energy $\epsilon_\mathrm{F}$:
$\epsilon_k=-2t \cos ka$ is replaced by 
$\epsilon_k=\hbar v_{\mathrm{F}} (|k|-k_{\mathrm{F}})$ 
with the Fermi wavenumber $k_{\mathrm{F}}$,
as shown in Fig.\ 2,  where $v_{\mathrm{F}}=(2ta/\hbar)
\sin k_{\mathrm{F}}a$. We assume that
$\epsilon_\mathrm{F} \simeq 0$
[$k_{\mathrm{F}} \simeq \pi/(2a)$] and set
the Fermi energy to be $\epsilon_{\mathrm{F}}=0$. 
Half of the bandwidth is
$D_0=\hbar v_{\mathrm{F}} k_{\mathrm{F}}$
($-2k_{\mathrm{F}}< k \le 2k_{\mathrm{F}}$).
The density of states in a lead is
constant;
$\rho(\epsilon_k)=Na/(\pi \hbar v_{\mathrm{F}})$, 
where $N$ is the number of sites in the lead.
This simplification is justified since the
wide-band limit is taken later.

For examining the Kondo effect, we focus on the Coulomb
blockade regime with one electron in the quantum dot,
which satisfies the conditions of $-\epsilon_0$,
$\epsilon_0+U \gg \Gamma$, $k_\mathrm{B}T$.
$\Gamma=\Gamma_L+\Gamma_R$ is the level broadening in the
quantum dot, where
$\Gamma_{\alpha}=\pi \nu_0 V_{\alpha}^2$ with 
$\nu_0=1/(\pi t)$ 
being the local density of states at the end of
semi-infinite leads at the Fermi level
$\epsilon_\mathrm{F}=0$. The background transmission
probability through the reference arm
is given by $T_\mathrm{b}=4x/(1+x)^2$ with $x=(W/t)^2$.

For comparison, we examine another model without the
reference arm:
a quantum dot with single energy level $\epsilon_0$ is
connected to two external leads by tunnel couplings,
$V_L$ and $V_R$, as shown in Fig.\ 1(b).

\begin{figure}
\includegraphics[width=9pc]{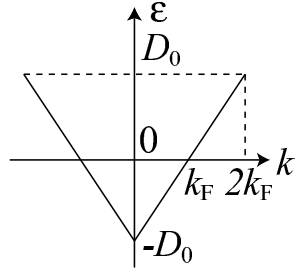}\hspace{1pc}%
\caption{
The energy dispersion in the external leads in Figs.\ 1(a)
and (b): $\epsilon_k=\hbar v_{\mathrm{F}} (|k|-k_{\mathrm{F}})$ 
for $-2k_{\mathrm{F}}< k \le 2k_{\mathrm{F}}$.
The half of the bandwidth is given by
$D_0=\hbar v_{\mathrm{F}} k_{\mathrm{F}}$.
}
\end{figure} 

\subsection{Equivalent model}

From the Hamiltonian in Eq.\ (\ref{eq:orig}), we construct
an equivalent model in which a quantum dot is coupled to
a single lead.
First, we diagonalize the Hamiltonian $H_\mathrm{leads+ring}$
for the outer region of the quantum dot.
There are two eigenstates for a given wavenumber $|k|$.
\begin{eqnarray}
| \psi_{k,\rightarrow} \rangle
& = &
\sum_{n \le 0} (e^{ikna}+r_k e^{-ikna}) |n \rangle
+ \sum_{n \ge 1} t_k e^{ik(n-1)a} e^{i\phi} |n \rangle,
\label{eq:right-going-wave}
\\
| \psi_{k,\leftarrow} \rangle
& = &
\sum_{n \le 0} t_k e^{-ikna} |n \rangle 
+ \sum_{n \ge 1} [e^{-ik(n-1)a}+r_ke^{ik(n-1)a}] e^{i\phi}
|n \rangle,
\label{eq:left-going-wave}
\end{eqnarray}
apart from a normalization factor, $1/\sqrt{2N}$. Here,
\begin{equation}
t_k=-\frac{\sqrt{x}(1-e^{2ika})e^{ika}}{1-xe^{2ika}}, \
r_k=-\frac{(1-x)e^{2ika}}{1-xe^{2ika}},
\end{equation}
and $| n \rangle$ is the Wannier function at site $n$.
$| \psi_{k,\rightarrow} \rangle$
($| \psi_{k,\leftarrow} \rangle$) represents the state
that an incident plane wave from the left (right) is
partly reflected to the left (right) and partly transmitted
to the right (left).
The spin index is omitted for now.
We perform a unitary transformation for these modes 
\begin{equation}
\left( \begin{array}{cc}
| \psi_{k} \rangle & | \bar{\psi}_{k} \rangle
\end{array}
\right)
=
\left( \begin{array}{cc}
| \psi_{k,\rightarrow} \rangle & | \psi_{k,\leftarrow} \rangle
\end{array}
\right)
\left( \begin{array}{cc}
A_k & -B_k^* \\ B_k & A_k^* 
\end{array}
\right),
\label{eq:unitary-trans}
\end{equation}
where $A_k$ and $B_k$ are determined such that
$\langle d | H_\mathrm{T} | \bar{\psi}_{k} \rangle=0$ 
with dot state $| d \rangle$. As a result,
mode $|\psi_k\rangle$ is coupled to the dot via $H_{\mathrm{T}}$,
whereas mode $|\bar\psi_k\rangle$ is completely decoupled. 

Neglecting the decoupled mode,
we obtain a model equivalent to the Hamiltonian in
Eq.\ (\ref{eq:orig}) in order to discuss the Kondo effect.
The tunnel coupling of $| \psi_{k} \rangle$ to the quantum dot
is described by
\begin{eqnarray}
|\langle d | H_\mathrm{T} | \psi_{k} \rangle |^2&=&
\frac{2V^2}{N}
\left\{1-\frac{1-x}{(1+x)^2}
\frac{1}{1-T_{\mathrm{b}}(\epsilon_k/D_0)^2}
[\cos 2k(l+1)a-x\cos 2kla]
\right. \nonumber\\
&&-2\left.\frac{\sqrt{\alpha x}\cos\phi}{(1+x)^2}
\frac{\sqrt{1-(\epsilon_k/D_0)^2}}{1-T_{\mathrm{b}}(\epsilon_k/D_0)^2}
[\sin 2k(l+1)a-x\sin 2kla]\right\}
\equiv |V_0(\epsilon_k)|^2,
\label{eq:VK0}
\end{eqnarray}
where $\alpha=4\Gamma_L \Gamma_R/(\Gamma_L+\Gamma_R)^2$ is the
asymmetric factor for the tunnel couplings of the quantum dot. 
We find that
\begin{eqnarray}
|V_0(\epsilon_k)|^2&=&
\frac{2V^2}{N}
\left( 1+\frac{1-x}{1+x}
\sin kL -\frac{2}{1+x}\sqrt{\alpha x}\cos\phi
\cos kL \right),
\label{VK}
\end{eqnarray}
in a wide-band limit, where half of the bandwidth
$D_0$ is much larger than $|\epsilon_0|$.
Since the strength of tunnel coupling
between the leads and dot is characterized by
$\rho(\epsilon_k) |V_0(\epsilon_k)|^2$, we can choose the
density of states in the lead, $\nu(\epsilon_k)$,
in such a way that
$\nu(\epsilon_k) V^2=\rho(\epsilon_k) |V_0(\epsilon_k)|^2$
with $V=\sqrt{V_L^2+V_R^2}$.
Then the Hamiltonian is written as
\begin{eqnarray}
H = 
\sum_{\sigma} \epsilon_0 d^\dagger_\sigma d_\sigma
+U \hat n_{\uparrow} \hat n_{\downarrow}
+\sum_{k,\sigma} \epsilon_k a^\dagger_{k,\sigma} a_{k,\sigma}
+\sum_{k,\sigma}V(d^\dagger_\sigma a_{k,\sigma}+\mathrm{h.c}.),
\label{eq:Hamiltonian}
\end{eqnarray}
with the density of states in the lead
\begin{equation}
\nu(\epsilon_k) =
\nu_0\left[
1+\sqrt{1-T_\mathrm{b}}
\sin\frac{\epsilon_k+D_0}{\epsilon_{\mathrm{T}}}
- P(\phi)\cos\frac{\epsilon_k+D_0}{\epsilon_{\mathrm{T}}} \right]
\label{eq:DOS}
\end{equation}
for $-D_0 \le \epsilon_k \le D_0$,
where $\epsilon_{\mathrm{T}}=\hbar v_{\mathrm{F}}/L$ is
the Thouless energy for ballistic systems,\cite{Altland}
or energy level spacing in an isolated ring.
Here,
\begin{equation}
P(\phi) =
\sqrt{\alpha T_\mathrm{b}} \cos \phi.
\label{eq:P}
\end{equation}
$|P(\phi)| \le 1$ since $0<\alpha$, $T_\mathrm{b} \le 1$.
All the interference effects in the ring, i.e.,
the AB oscillation and the
higher harmonics, are involved in the density of
states in Eq.\ (\ref{eq:DOS}).
It oscillates with the
period of $\epsilon_{\mathrm{T}}$, as schematically shown in
Fig.\ 3. We assume that $\epsilon_{\mathrm{T}} \ll D_0$.
The amplitude and phase of $\nu(\epsilon_k)$
depend on the magnetic flux
penetrating the ring through $P(\phi)$.

\begin{figure}
\includegraphics[width=9pc]{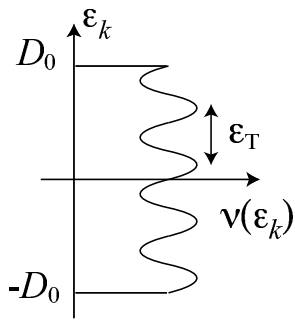}\hspace{1pc}%
\caption{
Schematic drawing of the density of states
$\nu(\epsilon_k)$ in a lead in the reduced model.
It oscillates with the period $\epsilon_{\mathrm{T}}$.
Its amplitude and phase depend on the magnetic flux penetrating
the ring through $P(\phi)$ in Eq.\ (\ref{eq:P}).} 
\end{figure}

\section{Case of $U\rightarrow \infty$}

The Hamiltonian (\ref{eq:Hamiltonian})
in the reduced model is analyzed by
the ``poor man's'' scaling method.\cite{Anderson}
We examine the case of $U \rightarrow \infty$ in this section.
The scaling procedure consists of two stages.\cite{Haldane}
On the first stage of the scaling, the charge fluctuation
is taken into account.
We reduce the energy scale from bandwidth $D_0$
until the charge fluctuation is quenched at $D=D_1$.
By integrating out the excitations in the energy range of
$D_1<D<D_0$, the energy level $\epsilon_0$ in the dot
is renormalized to $\tilde{\epsilon}_0$ ($D_1
\simeq |\tilde{\epsilon}_0|$).
On the second stage, we consider
the spin fluctuation at low energies of $D<D_1$.
We evaluate the Kondo temperature, using the Kondo
Hamiltonian.

In Appendix A, we illustrate the scaling procedure for the
model in Fig.\ 1(b) in which a quantum dot is connected to
two leads without the reference arm.
In its equivalent model, a quantum dot is coupled to a
lead, in which the tunnel coupling is $V=\sqrt{V_L^2+V_R^2}$
and the density of states in the lead is
$\nu(\epsilon_k)=\nu_0$.\cite{Glazman0}
The first stage scaling yields the renormalized energy level
\begin{equation} 
\tilde{\epsilon}_0^{(0)} \simeq
\epsilon_0+\nu_0 V^2\ln \frac{D_0}{|\epsilon_0|}.
\label{eq:level00}
\end{equation}
On the second stage, the Kondo temperature is
evaluated as
\begin{equation} 
T_{\mathrm{K}}^{(0)} \simeq
|\epsilon_0|\exp\left(
-\frac{1}{2\nu_0 J} \right),
\label{eq:TK00}
\end{equation}
where the exchange coupling is
$J=V^2/|\tilde{\epsilon}_0^{(0)}|$.

\subsection{Energy level renormalization}

Let us start the first stage of scaling using
the reduced model obtained in Sec.\ II.B.
The energy level in the quantum dot is evaluated by
$\epsilon_0=E_1-E_0$, where $E_0$ is the energy of
the empty state and $E_1$ is that of the singly
occupied state.
Reducing the bandwidth from $D$ to $D-|dD|$, they are
renormalized to $E_0+dE_0$ and $E_1+dE_1$, where
\begin{eqnarray*}
dE_0 & = &
-\frac{2V^2\nu(-D)}{D+E_1-E_0}|dD|,
\\
dE_1 & = &
-\frac{V^2\nu(D)}{D+E_0-E_1}|dD|,
\end{eqnarray*}
within the second-order perturbation with respect to tunnel
coupling $V$.
For $D \gg |E_1-E_0|$, they yield the scaling equation for
the energy level
\begin{equation}
\frac{d\epsilon_0}{d\ln D}=V^2
\left[\nu(D)-2\nu(-D)\right].
\label{eq:scaling0}
\end{equation}
Using the density of states $\nu(\epsilon_k)$ in
Eq.\ (\ref{eq:DOS}) and relation of
$D_0/\epsilon_{\mathrm{T}}=k_{\mathrm{F}}L$, we obtain
\begin{eqnarray}
\frac{d\epsilon_0}{d\ln D}
&=&-\nu_0 V^2 \left[
1+F_1(k_{\mathrm{F}}L,\phi)
\cos\frac{D}{\epsilon_{\mathrm{T}}}
+3F_2(k_{\mathrm{F}}L,\phi)
\sin\frac{D}{\epsilon_{\mathrm{T}}}
\right],
\label{eq:scaling01}
\end{eqnarray}
where
\begin{eqnarray}
F_1(k_{\mathrm{F}}L,\phi) &=&
\frac{\nu(\epsilon_{\mathrm{F}})-\nu_0}{\nu_0}
\nonumber \\
&=& \sqrt{1-T_\mathrm{b}}\sin k_{\mathrm{F}}L
-P(\phi)\cos k_{\mathrm{F}}L
\label{eq:F1}
\end{eqnarray}
and
\begin{equation}
F_2(k_{\mathrm{F}}L,\phi)=F_1(k_{\mathrm{F}}L+\pi/2,\phi).
\label{eq:F2}
\end{equation}
By the integration of Eq.\ (\ref{eq:scaling01}) from
$D_0$ to $D_1$, we obtain the renormalized energy level
\begin{eqnarray}
\tilde{\epsilon}_0
&=&\epsilon_0+\nu_0 V^2 \Biggl\{
\ln \frac{D_0}{D_1}-
F_1(k_{\mathrm{F}}L,\phi)
\left[\mathrm{Ci}\left(
\frac{D_1}{\epsilon_{\mathrm{T}}}\right)-
\mathrm{Ci}\left(
\frac{D_0}{\epsilon_{\mathrm{T}}}
\right)\right]
\nonumber\\
&& -3 F_2(k_{\mathrm{F}}L,\phi)
\left[
\mathrm{Si}\left(\frac{D_1}{\epsilon_{\mathrm{T}}}\right)
-\mathrm{Si}\left(\frac{D_0}{\epsilon_{\mathrm{T}}}\right)
\right]
\Biggr\},
\end{eqnarray}
where
\begin{eqnarray*}
\mathrm{Si}(x) \equiv \int^x_0 d\xi \frac{\sin{\xi}}{\xi},
\\
\mathrm{Ci}(x) \equiv \int_{-\infty}^x d\xi
\frac{\cos{\xi}}{\xi},
\end{eqnarray*}
and $D_1 \simeq |\tilde{\epsilon}_0|$.
Since $\Gamma=\pi \nu_0 V^2 \ll -\epsilon_0 \ll D_0$,
$D_1\simeq -\epsilon_0$. Thus
\begin{eqnarray}
\tilde{\epsilon}_0
&\simeq&
\epsilon_0+\nu_0 V^2 \left\{
\ln \frac{D_0}{|\epsilon_0|}
-F_1(k_{\mathrm{F}}L,\phi)
\mathrm{Ci}\left(
\frac{|\epsilon_0|}{\epsilon_{\mathrm{T}}}\right)
-3 F_2(k_{\mathrm{F}}L,\phi)\left[
\mathrm{Si}\left(
\frac{|\epsilon_0|}{\epsilon_{\mathrm{T}}}\right)
-\frac{\pi}{2}\right]
\right\}
\label{F0}
\end{eqnarray}
since $D_0\gg\epsilon_{\mathrm{T}}$. Here, we
have used asymptotic forms of 
$\mathrm{Si}(x)\sim {\pi}/{2}+{\cos x}/{x}$ and
$\mathrm{Ci}(x)\sim {\sin x}/{x}$ 
for $x\rightarrow\infty$.

From Eq.\ (\ref{F0}), we derive the renormalized level
in two situations,
(i) $|\epsilon_0| \gg \epsilon_{\mathrm{T}}$
and (ii) $|\epsilon_0| \ll \epsilon_{\mathrm{T}}$.
In situation (i), we find
\begin{equation}
\tilde{\epsilon}_0(\phi) \simeq \tilde{\epsilon}_0^{(0)},
\label{rne0}
\end{equation}
using the asymptotic forms of $\mathrm{Si}(x)$ and
$\mathrm{Ci}(x)$ at $x\rightarrow\infty$ again.
$\tilde{\epsilon}_0^{(0)}$ is given by Eq.\ (\ref{eq:level00}).
In this situation, the oscillating part of $\nu(\epsilon_k)$
in Eq.\ (\ref{eq:DOS}) is averaged out in the integration
of the scaling equation (\ref{eq:scaling01}). As a result,
the renormalization of energy level is not influenced by
the AB interference effect in the ring.

In situation (ii),
\begin{eqnarray}
\tilde{\epsilon}_0(\phi)&=&
\tilde{\epsilon}_0^{(0)}
-\nu_0 V^2\sqrt{1-T_\mathrm{b}}
\left[\frac{3\pi}{2}\cos k_{\mathrm{F}}L
+\left(\gamma+\ln \frac{|\epsilon_0|}{\epsilon_{\mathrm{T}}}\right)
\sin k_{\mathrm{F}}L\right]
\nonumber \\
&&-\nu_0 V^2 P(\phi)
\left[\frac{3\pi}{2}\sin k_{\mathrm{F}}L
-\left(\gamma+\ln \frac{|\epsilon_0|}
{\epsilon_{\mathrm{T}}}\right)\cos k_{\mathrm{F}}L\right],
\label{rne1}
\end{eqnarray}
using the other asymptotic forms of
$\mathrm{Si}(x)\sim x$ and $\mathrm{Ci}(x)\sim \gamma+\ln x$ 
for $x\rightarrow 0$. $\gamma\simeq 0.5772$ is the
Euler's constant. In this situation,
the renormalized level is modulated by the AB interference effect.

Conditions (i) and (ii) are rewritten as
$L \gg L_\mathrm{c}$ and $L \ll L_\mathrm{c}$,
respectively, where $L_\mathrm{c}=\hbar
v_\mathrm{F}/|\epsilon_0|$. $L_\mathrm{c}$ is the
screening length of charge fluctuation.
When $L \gg L_\mathrm{c}$, the screening of charge
fluctuation is hardly influenced by the AB interference
effect. Thus the renormalization of energy level is
independent of magnetic flux, as shown in Eq.\ (\ref{rne0}).
When $L \ll L_\mathrm{c}$, the screening is modulated
by $\phi$ and also changed by $k_{\mathrm{F}}L$,
following Eq.\ (\ref{rne1}).

\subsection{Evaluation of Kondo temperature}

On the second stage of scaling, we start from the
Hamiltonian (\ref{eq:Hamiltonian}) with renormalized energy
level $\tilde{\epsilon}_0$ and bandwidth
$D_1 \simeq |\epsilon_0|$.
To describe the spin fluctuation at the low-energy scale of
$D \ll D_1$, we make the Kondo Hamiltonian via the
Schrieffer-Wolff transformation
\begin{eqnarray}
H_\mathrm{Kondo} &=&
\sum_{k,\sigma}\epsilon_{k\sigma}a^\dagger_{k\sigma}a_{k\sigma}
+H_J+H_K,
\label{eq:kondo}
\\
H_J &=&
J\sum_{k^\prime,k}[S^+a^\dagger_{k^\prime\downarrow}a_{k\uparrow}
+S^-a^\dagger_{k^\prime\uparrow}a^\dagger_{k\downarrow}
 +S_z(a^\dagger_{k^\prime\uparrow}a_{k\uparrow}
     -a^\dagger_{k^\prime\downarrow}a_{k\downarrow})],
\\
H_K &=&
K\sum_{k^\prime,k}\sum_{\sigma}
a^\dagger_{k^\prime\sigma}a_{k\sigma},
\end{eqnarray}
where $S^+=d^\dagger_{\uparrow}d_{\downarrow}$,
$S^-=d^\dagger_{\downarrow}d_{\uparrow}$ and
$S_z=(d^\dagger_{\uparrow}d_{\uparrow}
    -d^\dagger_{\downarrow}d_{\downarrow})/2$ are the
spin operators in the quantum dot.
$H_J$ indicates the exchange coupling between the localized
spin and conduction electrons in the lead, whereas
$H_K$ represents the potential scattering of the conduction
electrons by the quantum dot. The coupling constants are 
\begin{equation}
J = \frac{V^2}{|\tilde{\epsilon}_0|}
\end{equation}
and
\begin{equation}
K = \frac{V^2}{2|\tilde{\epsilon}_0|}.
\end{equation}
Note that they depend on $\phi$ through
$\tilde{\epsilon}_0$ in Eq.\ (\ref{rne1}) in the
situation of $L \ll L_\mathrm{c}$, whereas
they do not in the situation of $L \gg L_\mathrm{c}$.
The density of states in the lead is given by
$\nu(\epsilon_k)$ in Eq.\ (\ref{eq:DOS}).

By changing the bandwidth, we renormalize the coupling constants
$J$ and $K$ so as not to change the low-energy physics
within the second-order perturbation with respect to
$H_J$ and $H_K$. The coupling constants follow the scaling equations
\begin{eqnarray}
\frac{dJ}{d\ln D}
&=&
-J^2\left[\nu(D)+\nu(-D)\right]-2JK\left[\nu(D)-\nu(-D)\right],
\label{eq:scaling2a}
\\
\frac{dK}{d\ln D}
&=&
-\left(\frac{3J^2}{4} +K^2\right)\left[\nu(D)-\nu(-D)\right].
\label{eq:scaling2b}
\end{eqnarray}
Using the density of states $\nu(\epsilon_k)$ in
Eq.\ (\ref{eq:DOS}), we obtain  
\begin{eqnarray}
\frac{dJ}{d\ln D}
&=&-2\nu_0 J^2
-2\nu_0 J^2 F_1(k_{\mathrm{F}}L,\phi)
\cos\frac{D}{\epsilon_{\mathrm{T}}}
+4\nu_0 JK F_2(k_{\mathrm{F}}L,\phi)
\sin\frac{D}{\epsilon_{\mathrm{T}}},
\label{eq:scalinga2}
\\
\frac{dK}{d\ln D}
&=&2\nu_0\left(\frac{3}{4}J^2+4K^2\right)
F_2(k_{\mathrm{F}}L,\phi)
\sin\frac{D}{\epsilon_{\mathrm{T}}},
\label{eq:scalingb2} 
\end{eqnarray}
where $F_1(k_{\mathrm{F}}L,\phi)$ and
$F_2(k_{\mathrm{F}}L,\phi)$ are given by Eqs.\
(\ref{eq:F1}) and (\ref{eq:F2}).
The energy scale $D$ where the fixed point of strong
coupling is reached determines the Kondo temperature.

We evaluate $T_\mathrm{K}$ in the following procedures.
First, scaling equations (\ref{eq:scalinga2}) and
(\ref{eq:scalingb2}) are analyzed in two extreme cases.
In the case of $D \gg \epsilon_{\mathrm{T}}$,
the oscillating part of the density of states
$\nu(\epsilon_k)$ is averaged out in the integration.
Then the scaling equations are effectively rewritten as
\begin{eqnarray}
\frac{dJ}{d\ln D} &\simeq & -2\nu_0 J^2,
\label{JlargeD}
\\
\frac{dK}{d\ln D} &\simeq& 0.
\label{KlargeD}
\end{eqnarray}
Thus the potential scattering is irrelevant to the
Kondo effect in this case.
In the case of $D\ll \epsilon_{\mathrm{T}}$,
the last term is much smaller than the other terms on the
right side of Eq.\ (\ref{eq:scalinga2}). Hence
the exchange coupling $J$ is renormalized by
\begin{equation}
\frac{dJ}{d\ln D}\simeq
-2\nu_0\left[1+ F_1(k_{\mathrm{F}}L,\phi)
\cos\frac{D}{\epsilon_{\mathrm{T}}}\right] J^2.
\label{JsmallD}
\end{equation}
The coupling constant $K$ is also renormalized although
its development is slower than that of $J$ by the factor of
$D/\epsilon_{\mathrm{T}}$. [As $D \rightarrow T_\mathrm{K}$,
$J$ and $K$ become $|K|/J=$constant ($\ll 1$) at the fixed
point of Eqs.\ (\ref{eq:scalinga2}) and (\ref{eq:scalingb2}),
as shown in Appendix C.]

Using Eqs.\ (\ref{JlargeD}) and (\ref{JsmallD}), we evaluate
the Kondo temperature in three situations,
(i) $\epsilon_{\mathrm{T}} \ll T_{\mathrm{K}}\ll |\epsilon_0|$,
(ii) $T_{\mathrm{K}} \ll \epsilon_{\mathrm{T}}\ll |\epsilon_0|$,
and
(iii) $T_{\mathrm{K}}\ll |\epsilon_0| \ll \epsilon_{\mathrm{T}}$.
The conditions correspond to
(i) $L_{\rm c} \ll L_{\rm K} \ll L$, 
(ii) $L_{\rm c} \ll L \ll L_{\rm K}$,
and (iii) $L \ll L_{\rm c} \ll L_{\rm K}$, respectively, where
$L_{\rm K}= v_{\rm F}\hbar/T_{\rm K}$ is
the screening length of spin fluctuation, i.e.\ size of the
Kondo screening cloud.

In situation (i), the scaling equation (\ref{JlargeD})
can be applicable until the scaling ends at
$D\simeq T_{\mathrm{K}}$, where $J \rightarrow \infty$.
By integrating the equation from
$D_1\simeq |\epsilon_0|$ to $T_{\mathrm{K}}$, we obtain
\begin{equation}
T_{\mathrm{K}}\simeq
|\epsilon_0|\exp\left(-\frac{1}{2\nu_0 J}\right).
\label{TKi}
\end{equation}
This is identical to $T_{\mathrm{K}}^{(0)}$ in Eq.\ (\ref{eq:TK00}).
The AB interference effect does not affect the energy-level
renormalization nor Kondo temperature when the ring size $L$
is much larger than both the screening length of charge
fluctuation $L_{\rm c}$ and that of spin fluctuation $L_{\rm K}$.

In situation (iii),
the scaling equation (\ref{JsmallD}) is valid in the whole
scaling region of $T_\mathrm{K}<D<D_1$, which yields
\begin{eqnarray}
T_{\mathrm{K}}\simeq |\epsilon_0|
\exp\left[-\frac{\chi(\phi)}{2\nu_0 J}\right],
\label{TKii}
\end{eqnarray}
where $\chi(\phi)=
\left[ 1+ F_1(k_{\mathrm{F}}L,\phi)\right]^{-1}=
\nu_0/\nu(\epsilon_{\mathrm{F}})$, or
\begin{equation}
\chi(\phi)=\left[
1+\sqrt{1-T_\mathrm{b}} \sin k_{\mathrm{F}}L
- P(\phi)\cos k_{\mathrm{F}}L \right]^{-1}.
\label{eq:chi}
\end{equation}
Using the renormlized energy in Eq.\ (\ref{rne1}),
we find
\begin{eqnarray}
T_{\mathrm{K}}(\phi)
&\simeq&
|\epsilon_0| \left(
\frac{T_{\mathrm{K}}^{(0)}}{|\epsilon_0|}
\right)^{\chi(\phi)
|\tilde\epsilon_0(\phi)/\tilde{\epsilon}_0^{(0)}|}.
\label{TKsmall}
\end{eqnarray}
In this situation, both $\tilde\epsilon_0(\phi)$
and $T_{\mathrm{K}}(\phi)$ are modulated by the magnetic flux
since $L$ is smaller than both the screening lengths.

In situation (ii), 
$T_{\mathrm{K}} \ll \epsilon_{\mathrm{T}} \ll |\epsilon_0|$.
The coupling constant $J$ is renormalized following
Eq.\ (\ref{JlargeD}) when $D$ is reduced
from $D_1$ to $\epsilon_{\mathrm{T}}$, and following
Eq.\ (\ref{JsmallD}) when $D$ is
reduced from $\epsilon_{\mathrm{T}}$ to $T_{\rm K}$.
We match the solutions of the respective 
equations aroud $D \simeq \epsilon_{\mathrm{T}}$ and obtain 
\begin{equation}
T_{\mathrm{K}}(\phi)
\simeq
\epsilon_{\mathrm{T}}e^{\gamma}
\left(
\frac{T_{\mathrm{K}}^{(0)}}{\epsilon_{\mathrm{T}}e^{\gamma}}
\right)^{\chi(\phi)}.
\label{TKiii}
\end{equation}
In this situation, the Kondo temperature reflects the AB
interference effect since the ring size $L$ is smaller than
the Kondo screening length $L_{\rm K}$, whereas the
energy-level renormalization does not since $L$ is larger
than $L_{\rm c}$.

The obtained results of $T_{\mathrm{K}}$ are plotted
in Fig.\ 4, as a function $\phi$, with (a) $k_{\mathrm{F}}L=0$
$\mathrm{mod}\ 2\pi$ and (b) $k_{\mathrm{F}}L=\pi$
$\mathrm{mod}\ 2\pi$. Since $T_{\mathrm{K}}(\phi)$ depends
on $\phi$ through $P(\phi)$ in Eq.\ (\ref{eq:P}),
it is a periodic function of $\phi$ and satisfies
$T_{\mathrm{K}}(\phi)=T_{\mathrm{K}}(-\phi)$.
$T_{\mathrm{K}}$ is significantly modulated by $\phi$
in situations (ii) and (iii), as shown by broken and solid
lines, respectively. In these situations,
$T_{\mathrm{K}}(\phi)$ is also changed by $k_\mathrm{F}L$
since the interference pattern is modified with $k_\mathrm{F}L$.
In situation (i),
$T_{\mathrm{K}}(\phi)=T_{\mathrm{K}}^{(0)}$,
irrespectively of $\phi$ and $k_\mathrm{F}L$ (dotted line).

\begin{figure}
\includegraphics[width=15pc]{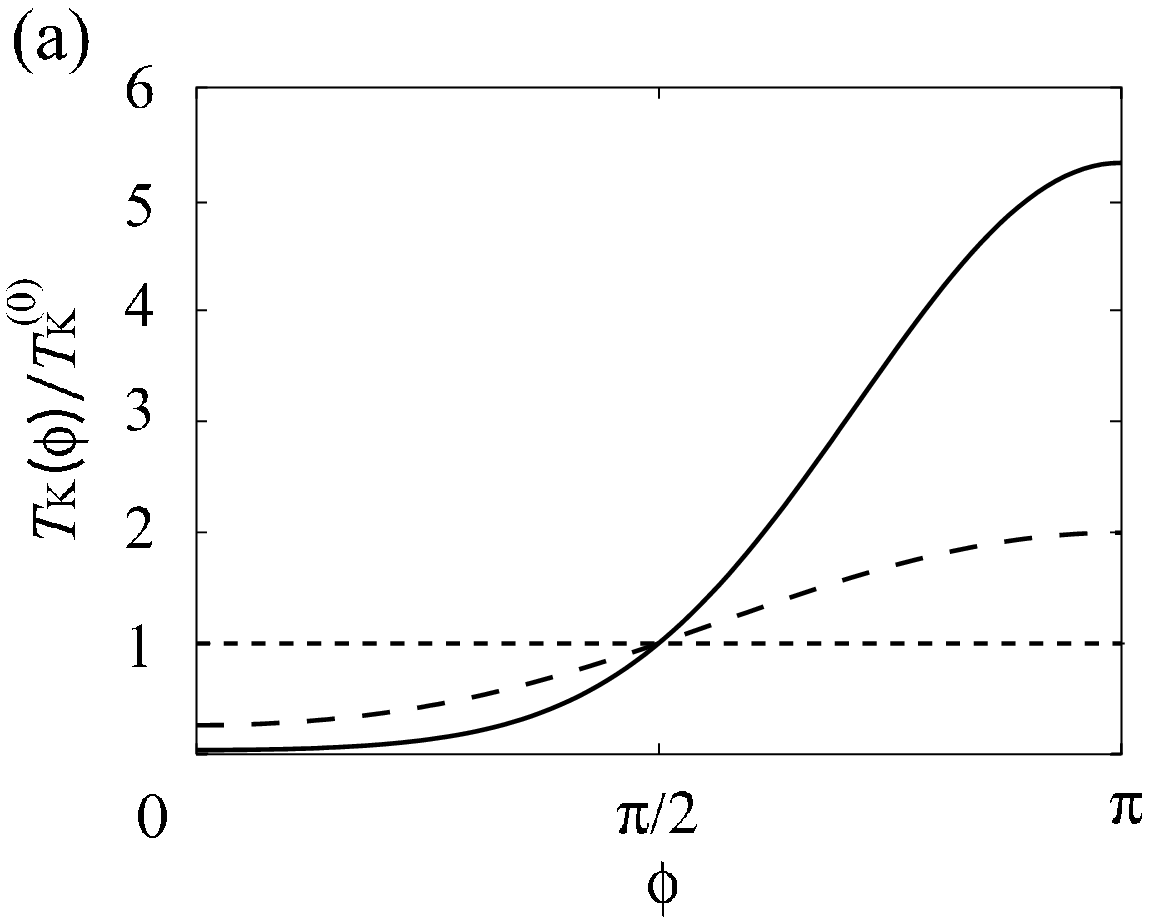}

\includegraphics[width=15pc]{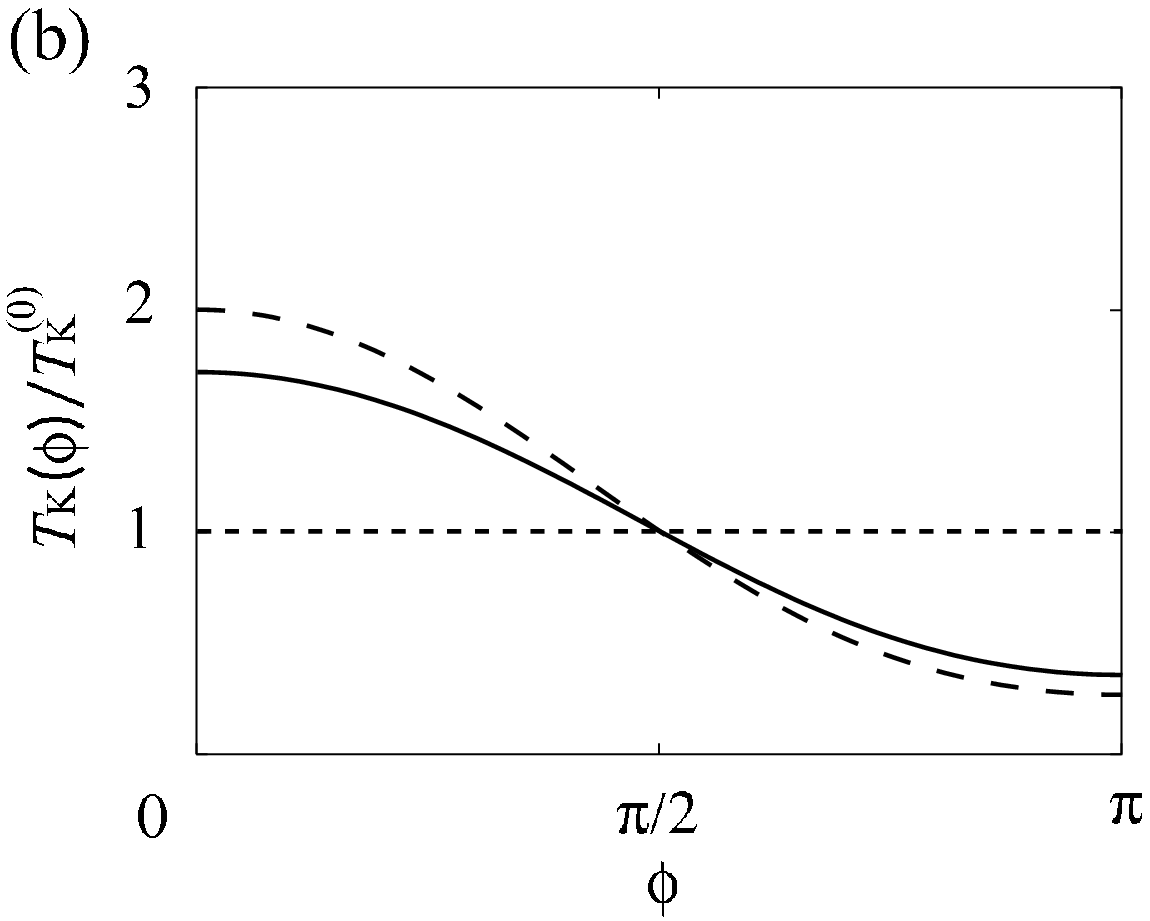}
\hspace{1pc}%
\caption{
The Kondo temperature $T_{\mathrm{K}}$ as a function of
AB phase $\phi$ of the magnetic flux penetrating the ring,
in the case of $U\rightarrow\infty$.
(a) $k_{\mathrm{F}}L=0$ $\mathrm{mod}\ 2\pi$
and (b) $\pi$ $\mathrm{mod}\ 2\pi$, where $L$ is the
ring size and $k_{\mathrm{F}}$ is the Fermi wavenumber.
$T_{\mathrm{K}}$ is normalized by
$T_{\mathrm{K}}^{(0)}$ in Eq.\ (\ref{eq:TK00}).
Situations (i) $L_{\rm c} \ll L_{\rm K} \ll L$, 
(ii) $L_{\rm c} \ll L \ll L_{\rm K}$,
and (iii) $L \ll L_{\rm c} \ll L_{\rm K}$ are
denoted by dotted, broken, and solid lines, respectively.
In situation (i), $T_{\mathrm{K}}=T_{\mathrm{K}}^{(0)}$,
irrespectively of the AB phase $\phi$.
}
\end{figure} 

\section{Vicinity of electron-hole symmetry}

In this section, we present the scaling analysis in the
vicinity of electron-hole symmetry;
$-\epsilon_0\simeq\epsilon_0+U$ (more precisely,
$|2\epsilon_0+U| <\Gamma$).
The scaling procedure is almost the same as in the
previous section. On the first stage of scaling, we
consider the energy level for the first electron
in the quantum dot, $\epsilon_0$, and that for the
second electron, $\epsilon_1=\epsilon_0+U$.

In Appendix A, the scaling analysis is given for the
model in Fig.\ 1(b) without the reference arm.
On the first stage, the energy levels are not
renormalized; 
\begin{equation} 
\tilde{\epsilon}_i^{(0)} \simeq \epsilon_i
\label{eq:level01}
\end{equation}
for $i=1,2$.
The second stage yields the Kondo temperature as
\begin{equation} 
T_{\mathrm{K}}^{(0)}=|\epsilon_0|
\exp\left( -\frac{1}{2\nu_0J} \right),
\label{eq:TK01}
\end{equation}
where $J={V^2}\left[
{|\epsilon_0|^{-1}}+(\epsilon_0+U)^{-1} \right]$.
Note that $T_{\mathrm{K}}^{(0)}$
in Eq.\ (\ref{eq:TK01}) is used in this section,
which is different from $T_{\mathrm{K}}^{(0)}$ in the
previous section [$J$ is not
identical in Eqs.\ (\ref{eq:TK00}) and (\ref{eq:TK01})].

\subsection{Energy level renormalization}

On the first stage of scaling, the charge fluctuation is
taken into account.
The energy levels in the quantum dot are given by
$\epsilon_0=E_1-E_0$ for the first electron and
$\epsilon_1=E_2-E_1$ for the second electron,
where $E_0$, $E_1$, and $E_2$ are
the energies of the empty state, singly occupied state,
and doubly occupied state in the quantum dot, respectively.
When the bandwidth is reduced from $D$ to $D-|dD|$,
$E_j$ ($j=0,1,2$) are renormalized to $E_j+dE_j$ with
\begin{eqnarray*}
dE_0 & = &
-\frac{2V^2\nu(-D)}{D+E_1-E_0}|dD|,
\\
dE_1 & = &
-\left[\frac{V^2\nu(D)}{D+E_0-E_1}+
\frac{V^2\nu(-D)}{D+E_2-E_1}\right]|dD|,
\\
dE_2 & = &
-\frac{2V^2\nu(D)}{D+E_1-E_2}|dD|,
\end{eqnarray*}
within the second-order perturbation with respect to tunnel
coupling $V$.
For $D \gg |E_1-E_0|$, $|E_2-E_1|$, they yield
the scaling equations for the energy levels
\begin{equation}
\frac{d\epsilon_0}{d\ln D} =
\frac{d\epsilon_1}{d\ln D} = 
2\nu_0 V^2 F_2(k_{\mathrm{F}}L,\phi)
\sin\frac{D}{\epsilon_{\mathrm{T}}},
\label{eq:scaling02}
\end{equation}
where $F_2(k_{\mathrm{F}}L,\phi)$ is given by Eq.\
(\ref{eq:F2}) in the previous section.

By the integration of Eq.\ (\ref{eq:scaling02}) from $D_0$
to $D_1$, we obtain the renormalized energy levels
\begin{eqnarray}
\tilde{\epsilon}_i
&=&\epsilon_i
+2\nu_0 V^2 F_2(k_{\mathrm{F}}L,\phi)
\left[
\mathrm{Si}\left(\frac{D_1}{\epsilon_{\mathrm{T}}}\right)
-\frac{\pi}{2}
\right],
\label{ehrnel}
\end{eqnarray}
for $i=0,1$, with $D_1 \simeq
\mathrm{max}(-\tilde{\epsilon}_0,\tilde{\epsilon}_1)$.
In the situation considered, $D_1\simeq-\tilde{\epsilon}_0
\simeq\tilde{\epsilon}_1$.

From Eq.\ (\ref{ehrnel}), we derive the renormalized level
in two situations,
(i) $|\epsilon_0| \gg \epsilon_{\mathrm{T}}$
and (ii) $|\epsilon_0| \ll \epsilon_{\mathrm{T}}$.
They correspond to (i) $L \gg L_\mathrm{c}$ and
(ii) $L \ll L_\mathrm{c}$, respectively, with
$L_\mathrm{c}=\hbar v_\mathrm{F}/|\epsilon_0|$ is the
screening length of charge fluctuation.
In situation (i), we find
\begin{eqnarray}
\tilde{\epsilon}_0
&\simeq& \epsilon_0,
\\
\tilde{\epsilon}_1
&\simeq& \epsilon_1=\epsilon_0+U.
\end{eqnarray}
The AB interference effect does not work on the level
renormalization since the ring size $L$ is larger than
the screening length of charge fluctuation.

In situation (ii),
the renormalized level is modulated by the AB interference
effect as
\begin{eqnarray}
\tilde{\epsilon}_i
&\simeq&\epsilon_i
-\pi\nu_0 V^2 F_2(k_{\mathrm{F}}L,\phi)
\nonumber \\
&=&
\epsilon_i
-\pi\nu_0 V^2
\left[ \sqrt{1-T_\mathrm{b}}\cos k_{\mathrm{F}}L
+P(\phi)\sin k_{\mathrm{F}}L \right].
\label{rneehp}
\end{eqnarray}

\subsection{Evaluation of Kondo temperature}

On the second stage, we derive the Kondo Hamiltonian
$H_\mathrm{Kondo}$ via the Schrieffer-Wolff transformation.
$H_\mathrm{Kondo}$ is in the same form as in Eq.\
(\ref{eq:kondo}) in the previous section, with
coupling constants of
\begin{eqnarray}
J&=&{V^2}\left(\frac{1}{|\tilde{\epsilon}_0|}
+\frac{1}{\tilde{\epsilon}_1}\right),
\label{coupling2}
\\
K&=&\frac{V^2}{2}\left(\frac{1}{|\tilde{\epsilon}_0|}
-\frac{1}{\tilde{\epsilon}_1}\right).
\end{eqnarray}
The energy levels $\tilde{\epsilon}_0$ and
$\tilde{\epsilon}_1$ are not renormalized when
$|\epsilon_0|\gg \epsilon_{\mathrm{T}}$, whereas
they are given by Eq.\ (\ref{rneehp})
when $|\epsilon_0| \ll \epsilon_{\mathrm{T}}$.
In both the situations, we find
\begin{equation}
J \simeq 
{V^2}\left( \frac{1}{|{\epsilon}_0|}+
\frac{1}{\epsilon_0+U} \right)
\label{Jehsym}
\end{equation}
to the order of $\Gamma/|\epsilon_0|$.

The scaling equations for $J$ and $K$ are derived within
the second-order perturbation with respect to
$H_J$ and $H_K$. They are identical to those in the
previous section, Eqs.\ (\ref{eq:scalinga2}) and
(\ref{eq:scalingb2}). From the equations, we obtain
Eq.\ (\ref{JlargeD}) in the case of
$D \gg \epsilon_{\mathrm{T}}$ and
Eq.\ (\ref{JsmallD}) in the case of
$D \ll \epsilon_{\mathrm{T}}$.

We evaluate the Kondo temperature
in three situations, (i) $L_{\rm c} \ll L_{\rm K} \ll L$, 
(ii) $L_{\rm c} \ll L \ll L_{\rm K}$,
and (iii) $L \ll L_{\rm c} \ll L_{\rm K}$.
In situation (i), $\epsilon_{\mathrm{T}}\ll T_{\mathrm{K}}
\ll |\epsilon_0|$.
The scaling equation (\ref{JlargeD}) yields
\begin{equation}
T_{\mathrm{K}}(\phi) \simeq T_{\mathrm{K}}^{(0)}
\end{equation}
which is the Kondo temperature of the model
in Fig.\ 1(b) without the reference arm [Eq.\ (\ref{eq:TK01})].
The AB interference effect is ineffective on the energy-level
renormalization and on the Kondo temperature.

In situation (iii),
$T_\mathrm{K} \ll |\epsilon_0| \ll \epsilon_{\mathrm{T}}$.
Then the scaling equation (\ref{JsmallD}) gives us
\begin{equation}
T_{\mathrm{K}}(\phi) \simeq 
|\epsilon_0|\left(
\frac{T_{\mathrm{K}}^{(0)}}{|\epsilon_0|}
\right)^{\chi(\phi)},
\label{eq:TKiii2}
\end{equation}
where the exponent $\chi(\phi)$ is given by Eq.\
(\ref{eq:chi}) in the previous section.
Since $L$ is smaller than $L_{\rm c}$ and $L_{\rm K}$,
both the energy levels and $T_{\mathrm{K}}(\phi)$ are
modulated by the magnetic flux. Because the exchange
coupling $J$ in Eq.\ (\ref{Jehsym}) is not influenced
by the energy-level renormalization, the expression
of $T_{\mathrm{K}}(\phi)$ is simpler than that in
Eq.\ (\ref{TKsmall}) in the previous section.

In situation (ii),
$T_\mathrm{K} \ll \epsilon_{\mathrm{T}} \ll |\epsilon_0|$.
$J$ is renormalized by Eq.\
(\ref{JlargeD}) at $\epsilon_{\mathrm{T}} \ll D \ll D_1
\simeq |\epsilon_0|$ and by Eq.\ (\ref{JsmallD}) at
$T_{\rm K} \ll D \ll \epsilon_{\mathrm{T}}$. We obtain
\begin{equation}
T_{\mathrm{K}}(\phi) \simeq 
\epsilon_{\mathrm{T}}e^\gamma
\left(\frac{T_{\mathrm{K}}^{(0)}}{
\epsilon_{\mathrm{T}}e^{\gamma}}\right)^{\chi(\phi)}.
\label{eq:TKii2}
\end{equation}

Figure 5 shows $T_{\mathrm{K}}(\phi)$
with (a) $k_{\mathrm{F}}L=0$
$\mathrm{mod}\ 2\pi$ and (b) $k_{\mathrm{F}}L=\pi$
$\mathrm{mod}\ 2\pi$. The behavior of $T_{\mathrm{K}}(\phi)$
is qualitatively the same as that in the previous
section with $U \rightarrow \infty$.
In situations (ii) and (iii),
$T_{\mathrm{K}}$ is significantly modulated by $\phi$
(broken and solid lines). It should be mentioned that
for a given $k_\mathrm{F}L$,
the modulation of $T_{\mathrm{K}}(\phi)$
is always larger in situation (iii) than in situation (ii)
in the vicinity of electron-hole symmetry
[Eqs.\ (\ref{eq:TKiii2}), (\ref{eq:TKii2})].
In situation (i), $T_{\mathrm{K}}=T_{\mathrm{K}}^{(0)}$,
irrespectively of $\phi$ and $k_\mathrm{F}L$ (dotted line).

\begin{figure}
\includegraphics[width=15pc]{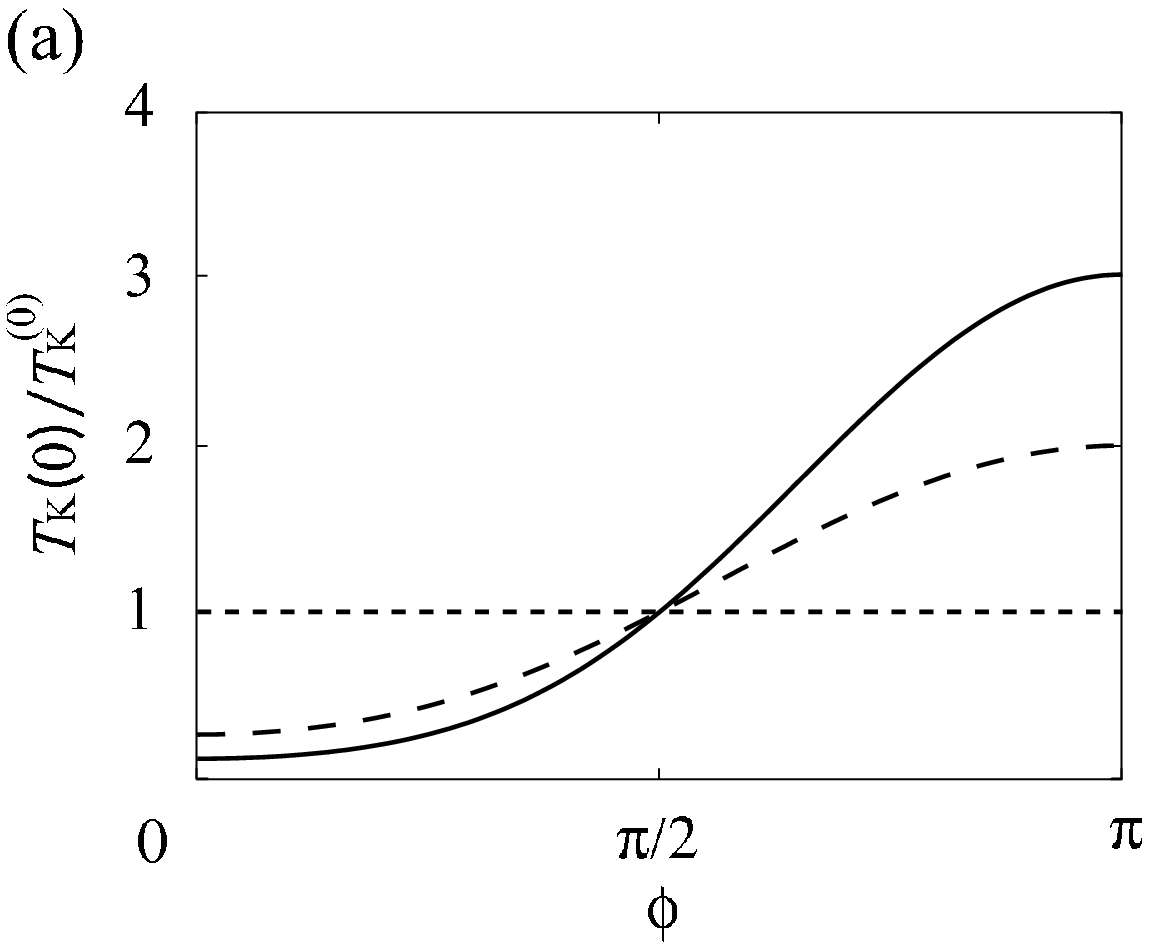}

\includegraphics[width=15pc]{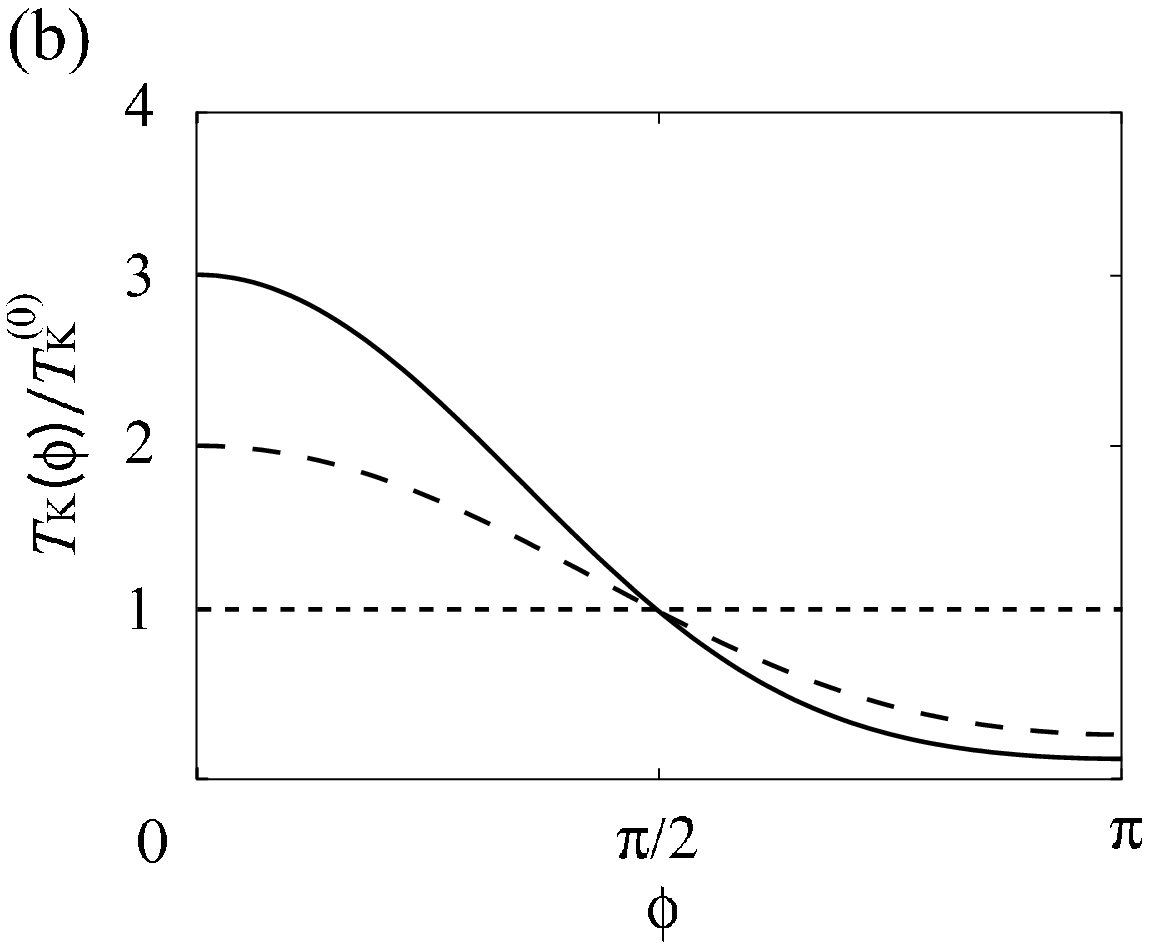}
\hspace{1pc}%
\caption{
The Kondo temperature $T_{\mathrm{K}}$ as a function of
AB phase $\phi$ of the magnetic flux penetrating the ring,
in the case of electron-hole symmetry,
$-\epsilon_0 = \epsilon_0+U$.
(a) $k_{\mathrm{F}}L=0$ $\mathrm{mod}\ 2\pi$
and (b) $\pi$ $\mathrm{mod}\ 2\pi$, where $L$ is the
ring size and $k_{\mathrm{F}}$ is the Fermi wavenumber.
$T_{\mathrm{K}}$ is normalized by
$T_{\mathrm{K}}^{(0)}$ in Eq.\ (\ref{eq:TK01}).
Situations (i) $L_{\rm c} \ll L_{\rm K} \ll L$, 
(ii) $L_{\rm c} \ll L \ll L_{\rm K}$,
and (iii) $L \ll L_{\rm c} \ll L_{\rm K}$ are
denoted by dotted, broken, and solid lines, respectively.
In situation (i), $T_{\mathrm{K}}=T_{\mathrm{K}}^{(0)}$,
irrespectively of the AB phase $\phi$.
}
\end{figure} 

\section{Conductance}

In this section, the conductance is evaluated on the basis
of the scaling analysis. First, we calculate the logarithmic
corrections
in the {\it weak-coupling} regime of $T \gg T_\mathrm{K}$.
The scattering of conduction electrons by a localized spin
in the dot is evaluated by the perturbation of $J$ in the
Kondo Hamiltonian.
Second, we obtain the analytical expression of the conductance
in the {\it strong-coupling} regime of $T \ll T_\mathrm{K}$.
We use the Hamiltonian in the strong-coupling fixed point
to describe the properties of the Fermi liquid.\cite{Glazman}
The calculations are applicable to both
the case of $U \rightarrow \infty$ and the vicinity of
electron-hole symmetry if $T_{\mathrm{K}}$ is replaced
by the value in respective cases.

\subsection{Weak-coupling regime}

At $T \gg T_\mathrm{K}$, a localized spin $1/2$ still remains
in the quantum dot. The scattering of conduction eletrons
by the localized spin can be
treated by the perturbation with respect to $J$ in the
Kondo Hamiltonian in Eq.\ (\ref{eq:kondo}).\cite{Hewson}
We solve a scattering problem for an incident wave of
$| \psi_{k,\rightarrow} \rangle$ in Eq.\
(\ref{eq:right-going-wave}) to the lowest order in $J$
(Born approximation).
When a localized spin in the quantum dot is in
the up-state, $| \mathrm{dot} \uparrow \rangle$,
an incident electron has an up- or down-spin,
$\sigma=\uparrow$, $\downarrow$.
The total wavefunction is written as
\begin{equation}
| \Psi_{\sigma} \rangle=
| \psi_{k,\rightarrow}; \sigma \rangle \otimes
| \mathrm{dot} \uparrow \rangle
+{\hat G}_0(\epsilon_k) H_J
\bigl(| \psi_{k,\rightarrow}; \sigma \rangle \otimes
| \mathrm{dot} \uparrow \rangle \bigr)
\label{eq:scattering}
\end{equation}
in the Born approximation, where ${\hat G}_0(\epsilon)$
is the unperturbed Green operator defined by
\begin{equation}
\left[\epsilon - H_\mathrm{leads+ring} +i\delta \right]
\hat G_0(\epsilon)=1.
\label{greeneq}
\end{equation}
Note that the first term in $H_\mathrm{Kondo}$ in
Eq.\ (\ref{eq:kondo}) is identical to
$H_\mathrm{leads+ring}$ in Eq.\ (\ref{eq:leads+ring})
if the decoupled modes are added.
We neglect the potential scattering $H_K$ in
$H_\mathrm{Kondo}$ since it is irrelevant to the Kondo
effect, or its logarithmic corrections are much smaller
than those of $H_J$ (see Sec.\ III.B).\cite{com1}
The matrix element of $H_J$ on the basis of the Wannier
function $| i \rangle$ is
given by $\langle i | H_J | j \rangle=
J (V_i V_j/V^2) {\bf S} \cdot {\bf s}$ ($i$, $j=$ $-l$ or
$l+1$), where $V_{-l}=V_L$, $V_{l+1}=V_R$,
${\bf S}$ is the spin operator in the quantum dot,
and ${\bf s}$ is that for a conduction electron.

For an incident electron with $\sigma=\uparrow$,
the localized spin remains in the up-state.
No spin-flip takes place.
The transmission probability is $|t_{\uparrow}|^2$,
where
\begin{eqnarray}
t_{\uparrow} &=&
\bigl(\langle l+1;\uparrow | \otimes
\langle \mathrm{dot} \uparrow |\bigr)
| \Psi_{\uparrow} \rangle
\nonumber
\\
&=&
\langle l+1| \psi_{k,\rightarrow} \rangle 
+\frac{1}{2V^2}\sum_{i,j=-l,l+1}V_iV_{j}
\langle l+1 |{\hat G}_0|i\rangle 
\langle j| \psi_{k,\rightarrow} \rangle.
\label{eq:t-up}
\end{eqnarray}
For an incident electron with $\sigma=\downarrow$,
there are two scattering processes in absence
($t_1$) or present ($t_2$) of the spin flip:
\begin{eqnarray}
t_{\downarrow,1} &=&
\bigl( \langle l+1;\downarrow | \otimes
\langle \mathrm{dot} \uparrow | \bigr)
| \Psi_{\downarrow} \rangle
\nonumber
\\
&=&
\langle l+1| \psi_{k,\rightarrow} \rangle 
-\frac{1}{2V^2}
\sum_{i,j=-l,l+1}V_iV_{j}\langle l+1 |{\hat G}_0|i\rangle 
\langle j| \psi_{k,\rightarrow} \rangle,
\label{eq:t-down1}
\\
t_{\downarrow,2} &=&
\bigl( \langle l+1;\uparrow | \otimes
\langle \mathrm{dot} \downarrow | \bigr)
| \Psi_{\downarrow} \rangle
\nonumber
\\
&=&
\frac{1}{V^2}
\sum_{i,j=-l,l+1}V_iV_{j}
\langle l+1 |{\hat G}_0|i\rangle 
\langle j| \psi_{k,\rightarrow} \rangle.
\label{eq:t-down2}
\end{eqnarray}
The transmission probability is given by
$|t_{\downarrow,1}|^2+|t_{\downarrow,2}|^2$.
The matrix elements of ${\hat G}_0(\epsilon_k)$,
$\langle l+1 |{\hat G}_0|-l\rangle$
and $\langle l+1 |{\hat G}_0|l+1\rangle$, are calculated
in Appendix D.
The conductance is evaluated by averaging over
the spin of incident electron $\sigma$:
\begin{equation}
G=\frac{e^2}{h} \left. (|t_{\uparrow}|^2
+|t_{\downarrow,1}|^2+ |t_{\downarrow,2}|^2)
\right|_{\epsilon_\mathrm{F}}.
\label{eq:weakGtmp}
\end{equation}

Finally, $J$ in Eq.\ (\ref{eq:weakGtmp}) is replaced by the
renormalized value at $D=k_\mathrm{B}T$, 
$\tilde{J}=[2\nu \ln(T/T_\mathrm{K})]^{-1}$. Then
we obtain the logarithmic corrections to the conductance
\begin{equation}
G_\mathrm{K}  = \frac{2e^2}{h}\frac{3\pi^2}
{16\left[\ln\left(T/T_{\mathrm{K}}\right)\right]^2}
\left[ T_{\mathrm{b}}R_{\mathrm{b}}
+\frac{\alpha}{4} \left(T_{\mathrm{b}}^2+R_{\mathrm{b}}^2
\right) +\sqrt{\alpha T_{\mathrm{b}}R_{\mathrm{b}}}
(T_{\mathrm{b}}+R_{\mathrm{b}})\cos\phi
+\frac{\alpha}{2} T_{\mathrm{b}}R_{\mathrm{b}}\cos 2\phi
\right],
\label{eq:G1}
\end{equation}
where
$R_{\mathrm{b}}=|\langle-l|\psi_{k,\rightarrow}\rangle|^2=
2-T_{\mathrm{b}}+2\sqrt{1-T_{\mathrm{b}}}\cos 2kla$.
In this approximation, the AB oscillation ($\cos{\phi}$;
due to the scattering from site $-l$ to $l+1$)
and second harmonics ($\cos{2\phi}$;
from site $l+1$ to $-l$) appear, whereas the
higher harmonics of the Fano resonance do not.
Note that $T_\mathrm{K}$ also changes with $\phi$ when
$L\ll L_{\mathrm{K}}$.

\subsection{Strong-coupling regime}

In the {\it strong-coupling} regime of  $T \ll T_\mathrm{K}$,
a spin $1/2$ in the quantum dot is fully screened out by the Kondo
effect. In this case, we can examine the scattering of
$| \psi_{k,\rightarrow} \rangle$ using the Fermi liquid theory.
For either spin $\sigma=\uparrow$ or $\downarrow$,
the scattered wave is written as
\begin{equation}
|\psi^\prime_k \rangle =
| \psi_{k,\rightarrow} \rangle 
+{\hat G}_0(\epsilon_k) \hat T
| \psi_{k,\rightarrow} \rangle,
\label{eq:scattering2}
\end{equation} 
where $\hat T$ is the t-matrix of the Kondo model in
Eq.\ (\ref{eq:kondo}) and $\hat{G}_0(\epsilon_k)$ is
the unperturbed Green operator in Eq.\ (\ref{greeneq}).
From Eq.\ (\ref{eq:unitary-trans}) in Sec.\ II.B, the incident
wave $| \psi_{k,\rightarrow} \rangle$ consists of two
parts:
\begin{equation}
| \psi_{k,\rightarrow} \rangle =A_k^\ast | \psi_{k} \rangle
- B_k | \bar{\psi}_{k} \rangle,
\end{equation}
where $| \psi_{k} \rangle$ is scattered by
$H_J + H_K$ while $| \bar{\psi}_{k} \rangle$ is not;
$\hat T|\bar\psi_k\rangle=0$.
If the potential scattering $H_K$ can be neglected,
$\langle \psi_k|\hat T|\psi_k \rangle \equiv T(\epsilon_k)$
is evaluated by the Hamiltonian in the strong coupling
fixed-point:\cite{Glazman} 
\begin{equation}
\pi \nu T(\epsilon) \simeq
\frac{\epsilon}{T_{\mathrm{K}}}
-i\left( 1-\frac{3\epsilon^2+\pi^2 T^2}{2T_{\mathrm{K}}^2} \right).
\label{t-matrix}
\end{equation}

The conductance $G$ is given by
\begin{equation}
G=\left.\frac{2e^2}{h}
|\langle l+1| \psi^{\prime}_k \rangle |^2
\right|_{\epsilon_k=\epsilon_{\mathrm{F}}},
\label{conductances}
\end{equation} 
where\cite{com3}
\begin{eqnarray}
\langle l+1 | \psi^{\prime}_k \rangle & = &
\langle l+1 | \psi_{k,\rightarrow} \rangle +
\sum_{k^\prime}
\langle l+1|{\hat G}_0(\epsilon_k) | \psi_{k^\prime} \rangle
\langle \psi_{k^\prime}|\hat T| \psi_k \rangle A_k^\ast
\nonumber \\
 & = & t_ke^{ikla}e^{i\phi}-i \pi \nu(\epsilon_k) T(\epsilon_k)
  \langle l+1 | \psi_k \rangle A_k^\ast.
\label{eq:tamp}
\end{eqnarray}
Using Eq.\ (\ref{t-matrix}) and
\begin{eqnarray}
\beta & \equiv &  \langle l+1 | \psi_k \rangle A_k^\ast
\nonumber \\
 & = & 
\frac{1}{2}
\frac{2t_ke^{i\phi} \left(1+r_ke^{2ikla}\right)
+\sqrt{\alpha} \left[t_k^2e^{2i\phi}e^{2ikla}+
\left(e^{-ikla}+r_ke^{ikla}\right)^2 \right]}
{1+\sqrt{1-T_b}\cos 2kla -P(\phi)\sin 2kla},
\end{eqnarray}
we obtain
\begin{equation}
G =\frac{2e^2}{h}
\left[ T_s+ (T_b-T_s)(\pi T/T_{\mathrm{K}})^2 \right],
\label{eq:G-strong-coupling}
\end{equation}
where $T_s=|t_ke^{ikla}e^{i\phi}-\beta|^2$. Although the
explicit expression of $T_s$ as a function of $\phi$ is
complicated in general, it is given in Appendix B for the
small limit of ring size ($l=0$). The same expression of $G$
can be obtained using the current formula in terms of the Green
function in the quantum dot.\cite{Hofstetter}

If the potential scattering $H_K$ is taken into account, the
deviation of $G$ from Eq.\ (\ref{eq:G-strong-coupling})
is expected to be very small in the vicinity of electron-hole
symmetry.\cite{Affleck4}
In the case of $U \rightarrow \infty$, on the other hand,
the deviation of $G$ might not be neglected,
as discussed by Yoshimori for the conventional Kondo
system of a magnetic impurity in metal.\cite{Yoshimori}

\section{Conclusions}

We have examined the Kondo effect in a quantum dot embedded
in a mesoscopic ring, using the ``poor man's'' scaling method.
For the purpose, we have constructed an equivalent model in which a
quantum dot is coupled to a single lead. The two-stage
scaling on the reduced model yields
analytical expressions of the Kondo
temperature $T_\mathrm{K}$ and conductance,
as a function of the Aharonov-Bohm phase $\phi$ by
the magnetic flux penetrating the ring. Regarding the
electron-electron interaction $U$ in the quantum dot,
we have examined both the cases of $U \rightarrow \infty$
and in the vicinity of electron-hole symmetry,
$-\epsilon_0 \simeq \epsilon_0+U$, where $\epsilon_0$
is the energy level in the quantum dot.

We have found two characteristic lengths in this Kondo
problem. One is the screening length of the charge
fluctuation, $L_\mathrm{c}=\hbar v_\mathrm{F}/|\epsilon_0|$,
and the other is
the screening length of spin fluctuation, i.e.,
size of Kondo screening cloud,
$L_\mathrm{K}=\hbar v_\mathrm{F}/T_\mathrm{K}$.
We obtain different expressions of $T_\mathrm{K}(\phi)$
for (i) $L_{\rm c} \ll L_{\rm K} \ll L$, 
(ii) $L_{\rm c} \ll L \ll L_{\rm K}$,
and (iii) $L \ll L_{\rm c} \ll L_{\rm K}$, concerning
the size of the ring $L$.
$T_\mathrm{K}$ is markedly modulated
by $\phi$ in cases (ii) and (iii), whereas it hardly depends
on $\phi$ in case (i). In the vicinity of electron-hole
symmetry, the modulation of $T_\mathrm{K}(\phi)$ with $\phi$
is larger in situation (iii) than in situation (ii).

We conclude the present paper with a few remarks. 
(i) We have restricted ourselves to the Kondo regime in
which the number of electrons is almost fixed
in the quantum dot.
When the energy level $\epsilon_0$ is tuned from the
Kondo regime to the
valence fluctuation regime by the gate voltage, this
system shows an asymmetric Fano-Kondo
resonance.\cite{Bulka,Hofstetter,Maruyama,Katsumoto}
Our scaling analysis, however, is applicable to the Kondo
regime only.
The examination of the crossover between the regimes
would require the calculations using the numerical
renormalization group,\cite{Hofstetter}
density-matrix renormalization group method,\cite{Maruyama}
etc., which is beyond the scope of the present paper.

(ii) In our model, the ring and external leads are
represented by one-dimensional tight-binding model,
in which the coherence is fully kept
for the AB interference effect. In a realistic case
with several conduction modes in the leads,
the coherence should be partly lost, as discussed
by Kubo {\it et al}.\ for a system of laterally
coupled double quantum dots.\cite{Kubo-Tokura}
Besides, in experimental results by Katsumoto
{\it et al}.,\cite{Katsumoto}
the conductance follows the Onsager's relation,
$G({\bf B})=G(-{\bf B})$, with magnetic field ${\bf B}$,
but does not satisfy $G(\phi)=G(-\phi)$ in some range
of magnetic field. This implies that the magnetic field
must be taken into account inside the arms of the ring
and leads as well as that penetrating the ring.
It requires the model with finite width of ring and leads.
Although the present model is insufficient for these
reasons, our calculation is straightforwardly generalized
to models with more than one mode in the ring and leads
with ${\bf B}$.

\section*{ACKNOWLEDGMENTS}

The authors acknowledge fruitful discussion with 
A.\ Aharony, O.\ Entin-Wohlman, I.\ Affleck, J.\ Malecki,
A.\ Oguri, and R.\ Sakano. This work was partly supported 
by a Grant-in-Aid for Scientific Research from the Japan
Society for the Promotion of Science, and by the Global COE
Program ``High-Level Global Cooperation for
Leading-Edge Platform on Access Space (C12).''

\appendix
\section{Scaling analysis of model in Fig.\ 1(b)}

To illustrate the two-stage scaling, we
perform the scaling analysis for the Kondo effect in a conventional
geometry depicted in Fig.\ 1(b): A quantum dot
with single energy level $\epsilon_0$ is connected to
two external leads by tunnel couplings, $V_L$ and $V_R$.

The eigenstates in leads $L$ and $R$ are given by
$| \psi_{k,\rightarrow} \rangle \equiv
| \psi_{k,L} \rangle$ and
$| \psi_{k,\leftarrow} \rangle \equiv
| \psi_{k,R} \rangle$, respectively, in Eqs.\
(\ref{eq:right-going-wave}) and
(\ref{eq:left-going-wave}) with $x=0$.
By the unitary transformation
\begin{equation}
\left( \begin{array}{cc}
| \psi_{k} \rangle & | \bar{\psi}_{k} \rangle
\end{array}
\right)
=
\left( \begin{array}{cc}
| \psi_{k,L} \rangle & | \psi_{k,R} \rangle
\end{array}
\right)
\left( \begin{array}{cc}
V_L/V & -V_R/V \\ V_R/V & V_L/V
\end{array}
\right),
\end{equation}
with $V=\sqrt{V_L^2+V_R^2}$, we obtain two modes;
mode $| \psi_{k} \rangle$ couples to the quantum dot
while mode $| \bar{\psi}_{k} \rangle$ is decoupled
from the dot.\cite{Glazman0}
Neglecting the latter, we obtain
the Hamiltonian in the same form as in Eq.\
(\ref{eq:Hamiltonian}) in the wide-band limit.
The density of states in the lead is
$\nu(\epsilon)=\nu_0$. 

Concerning the electron-electron interaction $U$ in
the quantum dot, we examine the case of
$U \rightarrow \infty$ first, and then in the
vicinity of the electron-hole symmetry,
$-\epsilon_0\simeq\epsilon_0+U$.

\subsection{Case of $U\rightarrow\infty$}

On the first stage of scaling,
we take into account the charge fluctuation and
renormalize the energy level $\epsilon_0$ to
$\tilde{\epsilon}_0$.
We reduce the energy scale from bandwidth $D_0$ to
$D_1$ where the charge fluctuation is quenched;
$D_1 \simeq -\tilde{\epsilon}_0$.

The energy level in the quantum dot is evaluated by
$\epsilon_0=E_1-E_0$, where $E_0$ is the energy of
the empty state and $E_1$ is that of the singly occupied
state. Reducing the bandwidth from $D$ to $D-|dD|$,
they are renormalized to $E_0+dE_0$ and $E_1+dE_1$, where
\begin{eqnarray}
dE_0 & = &
-\frac{2\nu_0 V^2}{D+E_1-E_0}|dD|,
\\
dE_1 & = &
-\frac{\nu_0 V^2}{D+E_0-E_1}|dD|,
\end{eqnarray}
within the second-order perturbation with respect to tunnel
coupling $V$.
For $D \gg |E_1-E_0|$, they yield the scaling equation for
the energy level
\begin{equation}
\frac{d\epsilon_0}{d\ln D}=-\nu_0 V^2.
\label{eq:0}
\end{equation}
By the integration of Eq.\ (\ref{eq:0}) from $D_0$ to $D_1$,
we obtain the renormalized energy level
\begin{eqnarray}
\tilde{\epsilon}_0^{(0)}
&=&\epsilon_0+\nu_0 V^2\ln \frac{D_0}{D_1},
\end{eqnarray}
where $D_1\simeq \tilde{\epsilon}_0^{(0)}$.
We put the superscript $(0)$ on the renormalized
level and Kondo temperature for the model in Fig.\ 1(b).
Since $\Gamma=\pi \nu_0 V^2 \ll -\epsilon_0 \ll D_0$ in
the Kondo regime, $D_1\simeq -\epsilon_0$. Therefore,
\begin{eqnarray}
\tilde{\epsilon}_0^{(0)}
&\simeq&
\epsilon_0+\nu_0 V^2\ln \frac{D_0}{|\epsilon_0|}.
\label{eq:1}
\end{eqnarray} 

On the second stage, we start from Hamiltonian
(\ref{eq:Hamiltonian}) with renormalized energy level 
$\tilde{\epsilon}_0^{(0)}$ and bandwidth
$D_1 \simeq |\epsilon_0|$. To take into consideration
the spin fluctuation at the low-energy
scale of $D \ll D_1$, we make the Kondo Hamiltonian
in Eq.\ (\ref{eq:kondo}) by the Schrieffer-Wolff
transformation.\cite{Hewson}
The coupling constants are 
\begin{eqnarray}
J &=& \frac{V^2}{|\tilde{\epsilon}_0^{(0)}|},
\\
K &=& \frac{V^2}{2|\tilde{\epsilon}_0^{(0)}|}.
\label{eq:2}  
\end{eqnarray}
By changing the bandwidth, we renormalize the coupling
constants $J$ and $K$ so as not to change the low-energy
physics within the second-order perturbation with respect to
$H_J$ and $H_K$. We obtain the scaling equations of
\begin{eqnarray}
\frac{dJ}{d\ln D}
&=&
-2\nu_0J^2,
\label{eq:scaling00}
\\
\frac{dK}{d\ln D}
&=& 0.
\label{eq:scaling00K}
\end{eqnarray}
Thus the potential scattering $K$ is irrelevant to the
Kondo effect. The energy scale $D$ where the fixed point
of strong coupling ($J\rightarrow \infty$) is reached
determines the Kondo temperature. The integration
of Eq.\ (\ref{eq:scaling00}) yields
\begin{equation}
T_{\mathrm{K}}^{(0)} \simeq
|\epsilon_0| \exp\left( -\frac{1}{2\nu_0 J} \right). 
\label{eq:3}
\end{equation}

\subsection{Vicinity of electron-hole symmetry}

In the case of $-\epsilon_0\simeq\epsilon_0+U$,
the number of the electrons in the quantum dot is
$0$, $1$, or $2$. We denote $E_0$, $E_1$, and $E_2$
for the energies of the empty state, singly occupied state,
and doubly occupied state in the quantum dot, respectively.
The energy levels in the quantum dot are
$\epsilon_0=E_1-E_0$ for the first electron and
$\epsilon_1=E_2-E_1$ for the second electron.
On the first stage of scaling, we renormalize them
by reducing the bandwidth from $D_0$ to $D_1$.

When the energy scale is changed from $D$ to $D-|dD|$,
$E_j$ ($j=0,1,2$) are changed to $E_j+dE_j$ with
\begin{eqnarray*}
dE_0 & = &
-\frac{2\nu_0V^2}{D+E_1-E_0}|dD|,
\\
dE_1 & = &
-\left[\frac{\nu_0V^2}{D+E_0-E_1}+
\frac{\nu_0V^2}{D+E_2-E_1}\right]|dD|,
\\
dE_2 & = &
-\frac{2\nu_0V^2}{D+E_1-E_2}|dD|.
\end{eqnarray*}
For $D \gg |E_1-E_0|$, $|E_2-E_1|$, they yield
the scaling equations for the energy levels
\begin{equation}
\frac{d\epsilon_0}{d\ln D} =
\frac{d\epsilon_1}{d\ln D} = 
0.
\end{equation}
Hence the energy levels are not renormalized:
\begin{eqnarray}
\tilde{\epsilon}_0^{(0)}
&=& \epsilon_0,
\\
\tilde{\epsilon}_1^{(0)}
&=& \epsilon_1=\epsilon_0+U.
\end{eqnarray}

On the second stage, we derive the Kondo Hamiltonian
in Eq.\ (\ref{eq:kondo}), as in the case of $U \rightarrow
\infty$. Now the coupling constants are 
\begin{eqnarray}
J &=& {V^2} \left(
\frac{1}{|\epsilon_0|} + \frac{1}{\epsilon_0+U}
\right),
\label{eq:J01}
\\
K &=& \frac{V^2}{2} \left(
\frac{1}{|\epsilon_0|} - \frac{1}{\epsilon_0+U}
\right).
\end{eqnarray}
$J$ and $K$ are renormalized following the scaling
equations (\ref{eq:scaling00}) and (\ref{eq:scaling00K}).
We obtain
\begin{equation}
T_{\mathrm{K}}^{(0)}=|\epsilon_0| \exp\left(
-\frac{1}{2\nu_0J}
\right)
\label{TK0ehsym}
\end{equation}
with $J$ in Eq.\ (\ref{eq:J01}).

\section{Scaling analysis in small limit of ring size}

In our previous paper,\cite{Yoshii}
we examined the model depicted in Fig.\ 1(a) 
in the small limit of ring size ($l=0$), using the same
method as in the present paper. Here,
we summarize the results for the following reasons.
(i) In Ref.\ \onlinecite{Yoshii}, we explained
the calculations in the case of $U\rightarrow\infty$
and presented the results only for the vicinity
of electron-hole symmetry. Now we showed the calculations
in both the cases.
(ii) Malecki and Affleck examined the same model and obtained
slightly different results from ours.\cite{Malecki}
The reason for the discrepancy is discussed.
(iii) The conductance at $T \ll T_\mathrm{K}$ is calculated
in two different ways. One is by the method given in Sec.\ V and
the other is using the current formula in terms of the Green
function in the quantum dot.\cite{Hofstetter} We obtain the identical results.

Note that the analytical expressions of the Kondo temperature
$T_\mathrm{K}(\phi)$ obtained in Sec.\ III and IV do not yield 
those in Eqs.\ (\ref{eq:TK-small-ring1}) and
(\ref{eq:TK-small-ring2}) in the limit of $L \rightarrow a$.
This is because the wide-band limit is taken for
Eq.\ (\ref{VK}), which is incompatible with the case of
$l=0$. On the other hand, the expressions of conductance
in Sec.\ V are applicable to the case of $l=0$ if
$T_\mathrm{K}(\phi)$ is given.

The Hamiltonian is given by Eq.\ (\ref{eq:orig}) with $l=0$.
First of all, we make an equivalent model in which a quantum dot
is coupled to a single lead, by the procedure presented in
Sec.\ II.A. The density of states in the lead is
\begin{equation}
\nu(\epsilon_k) =
\bar\nu\left[ 1+P(\phi)\frac{\epsilon_k}{D_0} \right],
\label{eq:DOS0}
\end{equation}
where $\bar\nu=\nu/(1+x)$. 
$P>0$ when $0 \le \phi < \pi/2$, whereas
$P<0$ when $\pi/2 < \phi \le \pi$. Thus $\nu(\epsilon_k)$
increases or decreases linearly with $\epsilon_k$,
in respective case, as shown in Fig.\ 6.
When $\phi=\pi/2$, $P=0$ and $\nu(\epsilon_k)$ is constant.

\begin{figure}
\includegraphics[width=25pc]{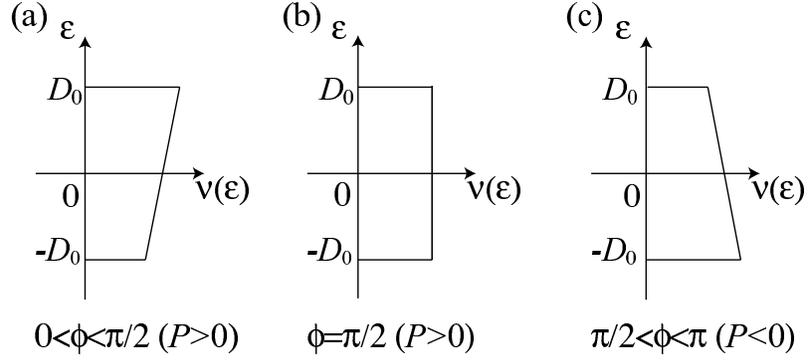}
\caption{
Schematic drawing of the density of states
$\nu(\epsilon_k)$ in Eq.\ (\ref{eq:DOS0})
in the reduced model in the case of small limit of ring
size [$l=0$ in Fig.\ 1(a)].
When $P(\phi)>0$ [$P(\phi)<0$], $\nu(\epsilon_k)$
increases [decreases] linearly with $\epsilon_k$.
}
\end{figure}

\subsection{Case of $U\rightarrow\infty$}

The case of $U\rightarrow\infty$ is examined in the similar
way. On the first stage, the scaling equation for $\epsilon_0$
is given by
\begin{equation}
\frac{d\epsilon_0}{d\ln D} =
-\bar\nu V^2
\left[ 1-3P_0(\phi)\frac{D}{D_0} \right].
\end{equation}
By the integration from $D_0$ to $D_1$ ($\simeq
-\epsilon_0$), we obtain
\begin{eqnarray}
\tilde{\epsilon}_0 & \simeq &
\epsilon_0+\bar\nu V^2
\left[\ln\frac{D_0}{|\epsilon_0|}-3P(\phi) \right]
\nonumber \\
& = &
\tilde{\epsilon}_0(\phi=\pi/2)-3 \bar\nu V^2 P(\phi).
\label{ep}
\end{eqnarray}

On the second stage of scaling, we make the Kondo
Hamiltonian in Eq.\ (\ref{eq:kondo}) with
$J={V^2}/{|\tilde{\epsilon}_0|}$ and
$K={V^2}/{2|\tilde{\epsilon}_0|}$.
The scaling equations for $J$ and $K$ are given by
Eqs.\ (\ref{eq:scaling2a}) and (\ref{eq:scaling2b}).
The substitution of $\nu(\epsilon_k)$ in Eq.\
(\ref{eq:DOS0}) yields
\begin{eqnarray}
\frac{dJ}{d\ln D}
&=&-2\bar\nu \left[ J^2-2JKP(\phi)\frac{D}{D_0} \right],
\label{eq:scalinga0}
\\
\frac{dK}{d\ln D}
&=&2\bar\nu\left(\frac{3}{4}J^2+4K^2\right)P(\phi)
\frac{D}{D_0}.
\label{eq:scalingb0}
\end{eqnarray}
As discussed in Ref.\ \onlinecite{Yoshii},
the second term in Eq.\ (\ref{eq:scalinga0}) can be
neglected. Equation (\ref{eq:scalinga0})
determines $T_\mathrm{K}$ as the energy scale $D$
at which $J$ diverges. It is
\begin{equation}
T_\mathrm{K}=D_1
\exp\left(
\frac{-1}{2\bar\nu J}
\right).
\label{kondo}
\end{equation}
By substituting $\tilde{\epsilon}_0$ in Eq.\
(\ref{ep}) into $J$, Eq.\ (\ref{kondo}) yields
\begin{equation}
T_\mathrm{K}(\phi)=
T_\mathrm{K} \left( \frac{\pi}{2} \right)
\exp\left[ -\frac{3}{2}P(\phi) \right].
\label{eq:TK-small-ring1}
\end{equation}
$T_\mathrm{K}(\phi)$ in Eq.\ (\ref{eq:TK-small-ring1}) is plotted by
solid line in Fig.\ 6. $T_\mathrm{K}(\phi)$ is
significantly modulated by the magnetic flux penetrating
the ring. $T_\mathrm{K}(\phi)$ is minimal at
$\phi=0$ and maximal at $\phi=\pi$ since the
renormalized level $\tilde{\epsilon}_0$ in Eq.\
(\ref{ep}) is the lowest (highest) at $\phi=0$
($\phi=\pi$).

\begin{figure}
\includegraphics[width=15pc]{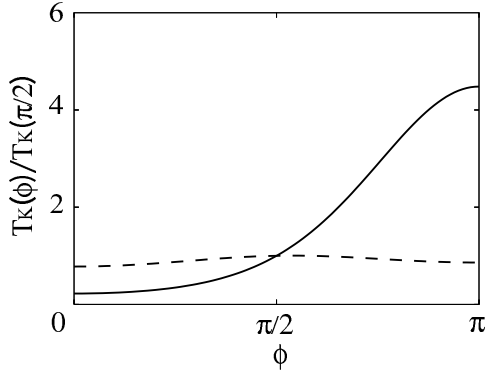}
\caption{
The Kondo temperature $T_{\mathrm{K}}$ as a function of
AB phase $\phi$ of the magnetic flux penetrating the ring,
in the small limit of ring size
[$l=0$ in Fig.\ 1(a)]. $T_{\mathrm{K}}(\phi)$
in the case of $U\rightarrow\infty$ is indicated by
solid line, whereas that in the vicinity of electron-hole
symmetry is indicated by broken line.
}
\end{figure} 

\subsection{Vicinity of electron-hole symmetry}

Next, we examine the vicinity of electron-hole symmetry.
On the first-stage scaling,
we renormalize the energy levels in the quantum dot,
$\epsilon_0$ for the first electron and
$\epsilon_1$ for the second electron.
They obey the scaling equations of
\begin{equation}
\frac{d\epsilon_0}{d\ln D} =
\frac{d\epsilon_1}{d\ln D} = 2\bar\nu V^2
P(\phi)\frac{D}{D_0}.
\label{eq:scaling12}
\end{equation}
By the integration of Eqs.\ (\ref{eq:scaling12})
from $D_0$ to $D_1$,
we obtain the renormalized energy level
\begin{equation}
\tilde{\epsilon}_i = 
\epsilon_i-2\bar\nu V^2
P(\phi)\left(1-\frac{D_1}{D_0}\right)
\end{equation}
for $i=1,2$, where $\epsilon_1=\epsilon_0+U$ and
$D_1 \simeq
\mathrm{max}(-\tilde{\epsilon}_0,\tilde{\epsilon}_1)$.
Since $\bar\nu V^2 \ll -\epsilon_0$,
$D_1\simeq -\epsilon_0\simeq \epsilon_1$ and thus 
\begin{eqnarray}
\tilde{\epsilon}_0 & \simeq &
\epsilon_0-2\bar\nu V^2 P(\phi),
\label{eq:level0}
\\
\tilde{\epsilon}_1 & \simeq &
\epsilon_1-2\bar\nu V^2 P(\phi).
\nonumber \\
& = & \epsilon_0+U-2\bar\nu V^2 P(\phi).
\label{eq:level1}
\end{eqnarray}

On the second stage, the coupling constants in the
Kondo Hamiltonian in Eq.\ (\ref{eq:kondo}) are
$J=V^2[|\tilde{\epsilon}_0|^{-1}
+(\tilde{\epsilon}_1)^{-1}]$ and
$K=(V^2/2)[|\tilde{\epsilon}_0|^{-1}
-(\tilde{\epsilon}_1)^{-1}]$. 
$J$ and $K$ are renormalized by Eqs.\
(\ref{eq:scalinga0}) and (\ref{eq:scalingb0}).
By neglecting the second term in Eq.\
(\ref{eq:scalinga0}), the Kondo temperature is
obtained as
\begin{equation}
T_\mathrm{K}(\phi) =
T_\mathrm{K} \left( \frac{\pi}{2} \right)
\exp\left[ -\frac{U+2\epsilon_0}{U}P(\phi)-
\frac{2\bar{\nu}V^2}{U} P(\phi)^2 \right].
\label{eq:TK-small-ring2}
\end{equation} 
We plot $T_\mathrm{K}(\phi)$ by broken line in
Fig.\ 7 when $-\epsilon_0=U+\epsilon_0$.
The $\phi$ dependence of $T_\mathrm{K}$ is very
weak compared with the case of $U \rightarrow \infty$.

The regime near the electron-hole symmetry was examined
by Malecki and Affleck.\cite{Malecki} They found that
$T_\mathrm{K}$ is independent of $\phi$ in the 
half-filling case of cosine band, $\epsilon_\mathrm{F}=0$, in
disagreement with Eq.\ (\ref{eq:TK-small-ring2}).
The $\phi$ dependence in Eq.\ (\ref{eq:TK-small-ring2})
stems from that of $\tilde{\epsilon}_0$ and
$\tilde{\epsilon}_1$ in Eqs.\ (\ref{eq:level0})
and (\ref{eq:level1}) which we obtain on the
first-stage scaling. In Ref.\ \onlinecite{Malecki},
the first-stage scaling was not performed:
(i) Malecki and Affleck made a reduced model by the same
method as in the present paper. In their reduced model,
a quantum dot with energy level $\epsilon_0$ is connected
to a lead with $\epsilon_k$-dependent tunnel
coupling $V(\epsilon_k)$
[see Eq.\ (\ref{eq:VK0}) in the present paper].
$V(\epsilon_k)$ usually depends on $\phi$, which  is
involved in the density of states in Eq.\ (\ref{eq:DOS0}) in our
treatment. For $\epsilon_k=0$, however, $V(\epsilon_k)$
is independent of $\phi$.
(ii) They performed the scaling of $J$
in the Kondo Hamiltonian, starting from
$J=V(\epsilon_\mathrm{F})^2
[|\epsilon_0|^{-1}+(\epsilon_0+U)^{-1}]$.
When $\epsilon_\mathrm{F}=0$,
they did not find the $\phi$-dependent $T_\mathrm{K}$
since $J$ is independent of $\phi$ .
This is inconsistent with our result in
Eq.\ (\ref{eq:TK-small-ring2}), which has been evaluated using
$\phi$-dependent $J$ through  $\tilde{\epsilon}_0$ and
$\tilde{\epsilon}_1$.
(iii) They observed the
$\phi$-dependent $T_\mathrm{K}$ in the case
of $\epsilon_\mathrm{F} \ne 0$, reflecting the
$\phi$-dependent $V(\epsilon_\mathrm{F})$ in $J$.

\subsection{Conductance}

The expressions of conductance $G$ obtained in Sec.\ V are
applicable to the case of $l=0$ if $T_{\mathrm{K}}$ is
given.
In the \textit{weak-coupling} regime of
$T\gg T_{\mathrm{K}}$, Eq.\ (\ref{eq:G1}) with $l=0$
yields the logarithmic corrections
\begin{eqnarray}
G_\mathrm{K} = \frac{2e^2}{h}\frac{3\pi^2}
{16\left[\ln\left(T/T_{\mathrm{K}}\right)\right]^2}
\left[
T_\mathrm{b}-2\sqrt{\alpha T_\mathrm{b}} \cos \phi
+ \alpha (1-T_\mathrm{b}+T_\mathrm{b}\cos^2 \phi)
\right].
\label{eq:G3}
\end{eqnarray}
In the {\it strong-coupling} regime of $T \ll T_\mathrm{K}$,
the analytical expression of $G$ is given by
Eq.\ (\ref{eq:G-strong-coupling}) with
\begin{equation}
T_s = \alpha(1-T_\mathrm{b} \cos^2\phi)
\label{eq:G4}
\end{equation}
if the potential scattering $H_K$ can be neglected.
This treatment should be justified in the vicinity of electron-hole
symmetry.\cite{Malecki,Affleck4}

We show an alternative method to derive the conductance at 
$T\ll T_{\mathrm{K}}$.\cite{Yoshii}
Following Ref.\ \onlinecite{Hofstetter},
the conductance is expressed in terms of Green function
in the quantum dot $G_\mathrm{dot}(\epsilon)$,
\begin{eqnarray}
G &=& \frac{2e^2}{h} \int d\epsilon
\left( -\frac{\partial f}{\partial \epsilon} \right) T(\epsilon),
\label{eq:cond1} \\
T(\epsilon) &=& T_\mathrm{b} +
2\sqrt{\alpha T_\mathrm{b}(1-T_\mathrm{b})}\cos\phi \bar{\Gamma}
{\rm Re}G_\mathrm{dot}(\epsilon)
\nonumber \\
& & 
-[\alpha(1-T_\mathrm{b} \cos^2\phi)-T_\mathrm{b}]
\bar{\Gamma} {\rm Im}G_\mathrm{dot}(\epsilon),
\label{eq:cond-Hofstetter}
\end{eqnarray}
where $f(\epsilon)$ is the Fermi distribution function and
$\bar{\Gamma}=\Gamma/(1+x)$.

$G_\mathrm{dot}(\epsilon)$ can be calculated using the
reduced model since the decoupled mode
from the quantum dot is not relevant. Using the Hamiltonian
in the strong coupling fixed-point,\cite{Glazman} we find
\begin{equation}
\bar{\Gamma} G_\mathrm{dot}(\epsilon)
\simeq
\frac{\epsilon}{T_\mathrm{K}}-
i \left(
1-\frac{3\epsilon^2+\pi^2 T^2}{2T_{\mathrm{K}}^2} \right).
\label{eq:Green-Hofstetter}
\end{equation}
The substitution of Eq.\ (\ref{eq:Green-Hofstetter}) into
Eq.\ (\ref{eq:cond-Hofstetter}) yields the conductance
in Eqs.\ (\ref{eq:G-strong-coupling}) and (\ref{eq:G4}).

\section{Fixed point of scaling equations
(\ref{eq:scaling2a}) and (\ref{eq:scaling2b})}

We show the fixed point of strong coupling in
the scaling equations (\ref{eq:scaling2a}) and
(\ref{eq:scaling2b}) when 
$T_\mathrm{K} \ll \epsilon_\mathrm{T}$, 
following the method
in Appendix A in Ref.\ \onlinecite{Eto}.
We introduce $y=K/J$ and $x=\ln(2\nu_0 J)$.
Clearly, $J \rightarrow \infty$ and thus
$x \rightarrow \infty$, as $D \rightarrow T_\mathrm{K}$.
For $y$, we obtain the equation of
\begin{equation}
\frac{dy}{dx}=-\frac{p(D) (y^2+\frac{3}{8}) + y}
{1 - p(D) y},
\label{sceqy}
\end{equation}
where
\begin{equation}
p(D)=\frac{2 F_2(k_{\mathrm{F}}L,\phi)
\sin\frac{D}{\epsilon_{\mathrm{T}}}}
{1+ F_1(k_{\mathrm{F}}L,\phi)
\cos\frac{D}{\epsilon_{\mathrm{T}}}}
\simeq 
\frac{2 F_2(k_{\mathrm{F}}L,\phi)}
{1+ F_1(k_{\mathrm{F}}L,\phi)}
\frac{D}{\epsilon_{\mathrm{T}}} 
\end{equation}
when $D \ll \epsilon_{\mathrm{T}}$.
From Eq.\ (\ref{sceqy}), we find a fixed point of
\begin{equation} 
y \simeq -\frac{3}{8}p(T_\mathrm{K})
\end{equation}
since $p(T_\mathrm{K})\ll 1$.
This means that $K$ and $J$ diverge simultaneously
at the fixed point, where
$y=K/J=O(T_\mathrm{K}/\epsilon_{\mathrm{T}}) \ll 1$.

\section{Matrix elements of Green function in Sec.\ V}

To calculate the conductance in the Born approximation,
we need some matrix elements of unperturbed Green
operator ${\hat G}_0(\epsilon_k)$.
From Eq.\ (\ref{greeneq}), we directly obtain the following
equations: For $n=-l$ or $l+1$ ($l \neq 0$),
\begin{equation}
\epsilon_k\langle m|{\hat G}_0|n\rangle = 
-t\langle m-1|{\hat G}_0|n\rangle
-t\langle m+1|{\hat G}_0|n\rangle
+\delta_{m,n}
\label{eq:grmn}
\end{equation}
for $m \neq 0,1$, and
\begin{eqnarray}
\epsilon_k\langle 0|{\hat G}_0|n\rangle &=& 
W e^{-i\phi}\langle 1|{\hat G}_0|n\rangle
-t\langle -1|{\hat G}_0|n\rangle,
\label{eq:gr0n}
\\
\epsilon_k\langle 1|{\hat G}_0|n\rangle &=& 
W e^{i\phi}\langle 0|{\hat G}_0|n\rangle
-t\langle 2|{\hat G}_0|n\rangle.
\label{eq:gr1n}
\end{eqnarray}
Since $\langle m|{\hat G}_0|-l\rangle$
describes the electron propagation from site $-l$ to $m$,
it can be expressed as
\begin{eqnarray}
\langle m|{\hat G}_0|-l\rangle &=&
Ae^{-ik(m+l)a}+Be^{-ik(m+l)a}
\hspace{1em} (m\leq -l),
\label{g-l1}
\\
\langle m|{\hat G}_0|-l\rangle &=&
Ae^{ik(m+l)a}+Be^{-ik(m+l)a}
\hspace{1em} (-l< m \leq 0),
\label{g-l2}
\\
\langle m|{\hat G}_0|-l\rangle &=&
Ce^{ik(m+l)a}e^{i\phi}
\hspace{1em} (m> 0),
\label{g-l3}
\end{eqnarray}
with unknown parameters, $A$, $B$, and $C$.
Similarly,
\begin{eqnarray}
\langle m|{\hat G}_0|l+1\rangle &=&
Ae^{ik(m-l-1)a}+Be^{ik(m-l-1)a}
\hspace{1em} (m> l+1),
\label{gl+11}
\\
\langle m|{\hat G}_0|l+1\rangle &=&
Ae^{-ik(m-l-1)a}+Be^{ik(m-l-1)a}
\hspace{1em} (1\leq m \leq l+1),
\label{gl+12}
\\
\langle m|{\hat G}_0|l+1\rangle &=&
Ce^{-ik(m-l-1)a}e^{-i\phi}
\hspace{1em} (m< 1).
\label{gl+13}
\end{eqnarray}
By substituting Eqs.\ (\ref{g-l1})--(\ref{gl+13}) and
$\epsilon_k=-2t\cos ka$ 
into Eqs.\ (\ref{eq:grmn})--(\ref{eq:gr1n}), 
we obtain
\begin{eqnarray*}
A &=&\frac{\pi\nu_0}{2i\sin ka},
\\
B &=&\frac{\pi\nu_0}{2i\sin ka}
\frac{(1-x)e^{ikL}}{xe^{ika}-e^{-ika}},
\\ 
C &=&-\pi\nu_0 \frac{\sqrt{x}e^{ikL}}{xe^{ika}-e^{-ika}},
\end{eqnarray*}
where $\nu_0=1/(\pi t)$ and $x=(W/t)^2$, as defined in Sec.\ II.

The matrix elements required in Sec.\ V are as follows.
\begin{eqnarray}
\langle l+1|{\hat G}_0|-l
\rangle &=& 
-\pi\nu_0\frac{\sqrt{x}e^{i(kL+\phi)}}{xe^{ika}-e^{-ika}},
\label{green1}
\\
\langle l+1|{\hat G}_0|l+1\rangle &=&
\frac{\pi\nu_0}{2i\sin ka}
\left[1+\frac{(1-x)e^{ikL}}{xe^{ika}-e^{-ika}}\right].
\label{green3}
\end{eqnarray}


\begin{thebibliography}{99}
\bibitem{com0}
We can choose the phase factors of Wannier functions
$\{ | n \rangle \}$ in the tight-binding model
and wavefunction $\{ | d \rangle \}$ in the quantum dot
in such a way that the AB phase appears only at the tunnel
barrier and $V_L$, $V_R>0$.
\bibitem{com1}
The cross terms between $H_J$ and $H_K$ disappear
when the conductance is averaged over the spin state
of an incident electron.
\bibitem{com3}
From Eq.\ (\ref{greeneq}),
$\langle \psi_{k^\prime}| \hat G_0(\epsilon_k) | \psi_{k^\prime}
\rangle =1/(\epsilon_k-\epsilon_{k^\prime}+i\delta)
=P/(\epsilon_k-\epsilon_{k^\prime})
-i \pi \delta(\epsilon_k-\epsilon_{k^\prime})$,
where $P$ represents the Cauchy principal value.
In Eq.\ (\ref{eq:tamp}), we replace the summation over
$k^\prime$ by the integral over $\epsilon_{k^\prime}$
and neglect the integral of
$P/(\epsilon_k-\epsilon_{k^\prime})$, considering
the wide-band limit.
\end{thebibliography}
\end{document}